\documentclass{emulateapj}
\pdfoutput=1

\usepackage{graphicx}
\usepackage{txfonts}
\usepackage{natbib}
\usepackage{hyperref}
\usepackage{ifpdf}
\usepackage{mathrsfs}
\long\def\beginpgfgraphicnamed#1#2\endpgfgraphicnamed{\resizebox{\hsize}{!}{\includegraphics{#1}}}

\ifpdf
    \DeclareGraphicsExtensions{.pdf,.png,.jpg}
\else
    \DeclareGraphicsExtensions{.eps}
\fi

\newcommand{\flash}{{\sc FLASH}}
\newcommand{\grad}[1]{\nabla #1}
\newcommand{\mvect}[1]{\mathbf{#1}}
\newcommand{\ud}{\mathrm{d}}
\newcommand{\Mdot}{\dot M}
\newcommand{\Msun}{\rm M_\odot}
\newcommand{\Rsun}{\rm R_\odot}
\newcommand{\yr}{\rm yr^{-1}}
\newcommand\kms{\ifmmode{\rm km\thinspace s^{-1}}\else km\thinspace s$^{-1}$\fi}
\newcommand\ms{\ifmmode{\rm m\thinspace s^{-1}}\else m\thinspace s$^{-1}$\fi}

\shorttitle{Numerical simulations of wind accretion}
\shortauthors{de Val-Borro, Karovska \& Sasselov}

\begin{document}

\title{Numerical Simulations of Wind Accretion in Symbiotic Binaries}

\author{M.~de~Val-Borro
\altaffilmark{1},
M.~Karovska
and D.~Sasselov}

\affil{Harvard-Smithsonian Center for
Astrophysics, 60 Garden St., Cambridge, MA 02138, USA}

\altaffiltext{1}{Stockholm University, AlbaNova
University Center, SE-106 91, Stockholm, Sweden}

\begin{abstract}
   {About half of 
the binary systems are close enough to
each other for mass to be exchanged between them at some
point in their evolution,
yet the
accretion mechanism in wind accreting binaries is not well
understood.}
   {We study the dynamical effects of gravitational
   focusing by a binary companion on winds from
   late-type stars.  In particular, we investigate the
   mass transfer and formation of accretion disks around
   the secondary in detached systems consisting of an
   AGB mass-losing star and an accreting companion.  The
   presence of mass outflows is studied as a function of
   mass loss rate, wind temperature and binary orbital
   parameters.}
   {A 2-dimensional hydrodynamical model is used to
   study the stability of mass transfer in wind
   accreting symbiotic binary systems.  In our
   simulations we use an adiabatic equation of state and
   a modified version of the isothermal
   approximation, where the temperature depends on the
   distance from the mass losing star and its companion.
   The code uses a block-structured adaptive mesh
   refinement method that allows us to have high
   resolution at the position of the secondary and
   resolve the formation of bow shocks and accretion disks.
   We explore the accretion flow between the components
   and formation of accretion disks for a range of
   orbital separations and wind parameters.}
   {Our results show the formation of stream flow
   between the stars and accretion disks of various
   sizes for certain orbital configurations.  For a typical
   slow and massive wind from an AGB star the flow
   pattern is similar to a Roche lobe overflow with
   accretion rates of 10\% of the mass loss from
   the primary. Stable disks with exponentially
   decreasing density profiles and masses of the order
   $10^{-4}$ solar masses are formed when wind acceleration
   occurs at several stellar radii.
   The disks are
   geometrically thin with eccentric streamlines and
   close to Keplerian velocity profiles.  The formation
   of tidal streams and accretion disks is found to be
   weakly dependent on the mass loss from the AGB star.}
   { Our simulations of gravitationally focused wind
   accretion in symbiotic binaries show the formation of
   stream flows and enhanced accretion rates onto the
   compact component.  We conclude that mass transfer
   through a focused wind is an important mechanism
   in wind accreting interacting binaries
   and can have significant impact on the evolution
   of the binary itself and the individual components.
   }
\end{abstract}


\section{Introduction}
\label{sec:intro}

Accretion is a very important energy source for a wide
variety of astronomical objects including many
interacting binaries.
The mass transfer in interacting binaries usually occurs 
via tidal interaction
and Roche lobe overflow (RLOF).  However, a wide range
of interacting binaries belong to a class of accreting
systems where the primary does not fill its Roche
surface, and the mass transfer is believed to occur
mostly by wind accretion.  These include many symbiotic
binaries and even more exotic systems such as massive
X-ray binaries and microquasars.  Wind accretion is more
complicated than tidal interaction, and the accretion
processes in wind accreting binaries are poorly
understood.  
In this paper, we study
gravitationally focused wind accretion in symbiotic systems with an 
AGB star and
a compact accretor where the photosphere of the primary
does not fill its Roche lobe.

Symbiotic binaries are important astrophysical
laboratories for studies of wind
accretion because of the wide separation of the components,
and the ability to study the individual components and the accretion
processes
at many wavelengths ranging from X-ray to radio
\citep{1997ApJ...482L.175K,2005ApJ...623L.137K,2006ApJ...637L..49M}.
A typical symbiotic
consists of a mass-losing AGB or a red giant star and a
hot accreting companion, often a white dwarf (WD). 
The components in these systems are assumed to be detached 
(at least during most of
the orbital motion) and the compact companion accretes
mass from the massive wind of the cool evolved star.

Mass loss is known to be a key factor in the late stages
of the evolution of an AGB star.  
Matter escapes easily
because of the low surface gravity.
As the intermediate mass stars enter the AGB phase their atmospheres rise to
sizes of the order of few AU.
These stars become unstable to large
amplitude radial pulsations and drive strong shock waves
that produce an extended envelope and a copious
stellar wind driven by radiation pressure on dust grains
that form at several stellar radii above the photosphere
\citep{1988ApJ...329..299B}.
The powerful stellar winds also affect the evolution
of the ISM by influencing its dynamics and spreading
elements generated in stellar interiors.  It is believed
that interstellar dust has its origin in cool
atmospheres and dense winds blowing from these evolved
stars.

Due to massive stellar winds, intermediate mass stars
can avoid exploding as a supernova and become planetary nebulae (PN)
and proto-planetary nebulae (PPN)
at the end of the AGB
\citep[see e.g.][]{2002A&A...385..205G}.
Mass-losing giant stars and PPNs often present asymmetric
envelopes and bipolar outflows that can be explained by wind gravitational
focusing and wind collision dynamics.
The presence of bipolar and collimated outflows in PPNs
has been explained due to a collision of stellar winds
\citep{1997MNRAS.292..795M,2004ApJ...600L..59G}
and magnetically driven jets \citep{2001Natur.409..485B}.

In the following we show that
although the photosphere of the AGB star
in symbiotic systems  does not fill its Roche lobe,
its extended atmosphere can be
significantly disturbed by the accreting companion.  
Mass accretion 
can be increased due to the gravitational
attraction from the accretor and the centrifugal force
due to the rotation of the system. 
This is important, because 
the binary parameters
can  evolve which will affect the stellar
mass loss and wind parameters.  Mass and angular
momentum can be lost from the system or accreted onto
the companion.
The angular momentum loss can result in
orbital shrinking of the system which would enhance mass
transfer and can lead to a Type Ia supernova
\citep{1999ApJ...522..487H,2005A&A...441..589J,2007BaltA..16...26P}.
The evolution of AGB stars in binary systems
gives rise to highly aspherical winds and
can be significantly different from that of single
stars.

We use 2-dimensional hydrodynamical
time-dependent models to investigate the mass transfer
driven by a massive AGB wind that is captured by a detached
companion star.
We calculate accretion rates assuming wind acceleration
at different radii of the extended atmosphere
(dust formation radii)
and also consider the case
when the extended envelope reaches the Lagrangian point $L_1$
and the gas flows in a narrow stream towards
the secondary.
In Section~\ref{sec:wind}, we discuss
Bondy-Hoyle wind accretion in a symbiotic systems with an AGB
mass-losing star.  We describe our numerical setups in
Section~\ref{sec:model}.  In Section~\ref{sec:results}
we present the results of the numerical simulations and
we compare the theoretical models with recent
observations of symbiotic binaries to improve our
understanding of the properties of these systems.
Finally, we discuss the results and future work
in Section~\ref{sec:discuss}.

\section{Bondi-Hoyle accretion}
\label{sec:wind}

Physical processes
involved in winds from extended atmospheres are very
complex and only recently a detailed theoretical
modeling has been possible \citep[see
e.g.][]{1991ApJ...375L..53B,1997A&A...327..614A,
1998ApJ...497..303M,
2004A&A...413..789S,2008A&A...483..571F}.
In the classical theory of plane-parallel stellar
atmospheres the physical structure of the atmosphere is
determined by the effective temperature and surface
gravity for a given chemical composition of the
atmosphere.  In the case of the AGB stars the situation
becomes more complex. Pulsations in the atmosphere,
phenomena generated in the stellar interior such as
convection, and especially dust formation  influence the mass 
loss and wind parameters.

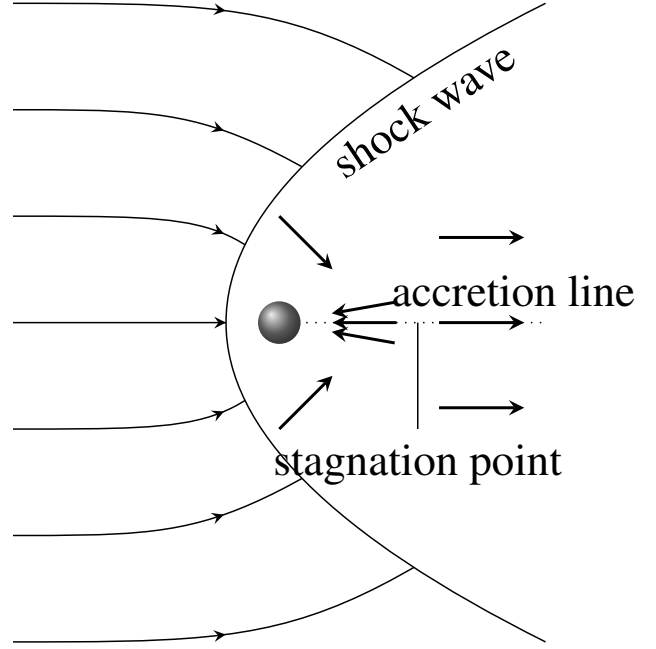
\begin{figure}
    \centering
    \beginpgfgraphicnamed{f01}
    \begin{tikzpicture}[decoration={ markings,
    mark=at position 2cm with {\arrow[black]{stealth}}} ]
        \tikzstyle{post}=[->,>=stealth,thick]
        \draw (5,3) ..controls (1,1) and (1,-1) ..(5,-3)
        node [very near start,sloped,below] {shock wave};
        \draw[post] (3.6,0) -- (3,0);
        \draw[post] (4.,0) -- (4.8,0);
        \draw[dotted] (2.5,0) -- (5,0);
        \shade[ball color=gray] (2.5,0) circle (.2);
        \draw (3.8,0) -- (3.8,-1) node [below] {stagnation point};
        \draw (4.7,0) node [above] {accretion line};
        \foreach \y in {.8} {
            \foreach \s in {+,-} {
                \draw[post] (4.,\s\y) -- (4.8,\s\y);
        }}
        \draw[post] (2.5,1) -- (3,.5);
        \draw[post] (2.5,-1) -- (3,-.5);
        \begin{scope}[xshift=2.5cm]
            \foreach \y in {+,-}
            {
               \draw[post] (\y10:1.1) -- (\y10:.5);
            }
        \end{scope}
        \clip (0,-3.1) -- (5,-3) ..controls (1,-1) and
        (1,1) ..(5,3) -- (0,3.1);
        \draw[postaction={decorate}] (0,0) -- (3,0);
        \foreach \s in {+,-} {
            \draw[postaction={decorate}] (0,\s1) ..controls (2,\s1) ..(3,0);
            \draw[postaction={decorate}] (0,\s2) ..controls (2,\s2) ..(5,0);
            \draw[postaction={decorate}] (0,\s3) ..controls (3,\s3) ..(7,0);
        }
    \end{tikzpicture}
    \endpgfgraphicnamed
    \caption[Bondi-Hoyle accretion geometry]
    {Diagram of the accretion geometry around the
    secondary marked by the gray circle.
    The flow is deflected at the shock wave
    and most of the accretion takes place
    along the accretion line behind the accretor.}
    \label{fig:bondi}
\end{figure}

For high velocity winds where the
pressure can be neglected, the Bondi-Hoyle accretion
rate is given by
\citep[see][for a recent review of Bondi-Hoyle-Lyttleton
accretion]{2004NewAR..48..843E}
\begin{equation}
    {\dot M}_{\rm BH} = \pi R_{\rm a}^2 \rho_{\rm B} v_{\rm r},
\end{equation}
where
$\rho_{\rm B}$ is  density at the secondary's position,
$v_{\rm r} = \sqrt{v_{\rm w}^2+v_{\rm B}^2}$
is the wind velocity 
relative to the accreting object,
$v_{\rm w}$ is the wind velocity at the position of the secondary,
$v_{\rm B}$ is the orbital velocity of the companion,
and $R_{\rm a}$ is the Bondi-Hoyle accretion radius
given by
\begin{equation}
    R_{\rm a} = \frac{2\, GM_{\rm B}}{v_{\rm r}^2}.
\end{equation}
When the mass accretion rate is normalized to the
2-dimensional case we obtain
\begin{equation}
    {\dot M}_{\rm BH} = 2 R_{\rm a} \rho_{\rm B} v_{\rm r} .
    \label{eq:accret}
\end{equation}

The mass loss from the unperturbed wind of the AGB
component in the orbital plane at the secondary position is
\begin{equation}
    {\dot M}_{\mathrm wind} = 2 \pi d \rho_{\rm B} v_{\rm w}.
\end{equation}
Therefore the Bondi-Hoyle accretion ratio, defined as
the mass accretion divided by the mass loss rate from
the primary, in our 2-dimensional model is given by
\begin{equation}
    f_{\rm BH} = \frac{{\dot M}_{\rm BH}}{{\dot M}_{\rm
    wind}}
    = \frac{\Omega_{\rm b}^2 d^2}{\pi v_{\rm r} v_{\rm w} (1+q)},
\end{equation}
with $q$ denoting the mass ratio, $d$ the
binary separation and we have used the definition of the orbital
Keplerian frequency
\begin{equation}
   \Omega_{\mathrm{b}} =  \left( \frac{G
   (M_{\mathrm{A}} + M_{\mathrm{B}})}{d^3} \right)^{1/2}.
   \label{eq:kepler}
\end{equation}
As we will see in Section~\ref{sec:resultswind}, the fraction of the AGB
wind accreted by the secondary
is several times higher than the Bondi-Hoyle rate
in our simulations due to gravitational wind focusing.
However, the Bondi-Hoyle theory cannot
be directly applied to the wind accretion scenario
in the case of slow winds where the sound speed is not
negligible.

If the gas pressure is included
in the Bondi-Hoyle scenario
\citep{1952MNRAS.112..195B,1992ApJ...384..587T}, the mass accretion ratio 
in the orbital plane becomes
\begin{equation}
    f_{\rm BH} = \frac{\Omega_{\rm b}^2 d^2}{\pi v_{\rm
    r} v_{\rm w} (1+q)}
    \left(\frac{\mathscr{M}}{1+\mathscr{M}^2}\right)^{3/2},
\end{equation}
where $\mathscr{M}$ is the Mach number
\begin{equation}
    \mathscr{M} = \frac{v_\mathrm{w}}{c_\mathrm{s}},
\end{equation}
and $c_{\mathrm s}$is the sound speed.
The accretion radius is defined as
\begin{equation}
    R_{\rm a} = \frac{2\, GM_{\rm B}}{v_{\rm r}^2+c_{\rm{s}}^2}.
\end{equation}

A diagram of the flow structure in
the orbital plane is shown in Fig.~\ref{fig:bondi}.
The stellar wind flows from the left and impacts
the shock wave and is deflected towards the accreting
companion. In the Bondi-Hoyle picture, the gas 
reaches the accretion line where the transversal
velocity component is neglected and the gravitationally
bounded material is accreted.
In our wind accretion model the AGB photosphere does not fill
its Roche lobe and the mass loss is carried out in the form of a
slow wind from the
region where dust grains are condensed.

The wind in our models is characterized
by a typical AGB mass-loss rate $\Mdot \sim
10^{-9}$--$10^{-6}~\Msun~\yr$
\citep{2000ApJ...545..945R} and the unperturbed
initial velocity profile is given by the isothermal wind
solution.
In our slower wind models the Bondi accretion radius is
comparable to the stellar separation and Bondi-Hoyle
theory is not valid. Nevertheless, we will use the
Bondi-Hoyle rates to compare with the mass accretion
rates resulting from our simulations.

\section{Numerical model}
\label{sec:model}

Two different problems in the context of mass loss in
late stages of stellar evolution are investigated.  In
the first case the primary's extended envelope fills its
Roche lobe and the gas flows in a narrow stream towards
the secondary. This could be the case when the companion is in
periastron.
We also use this model to test our numerical model.
Mass transfer in the second model is
driven by a massive wind that is captured by a detached
companion star.  This is an important mechanism for the
formation of accretion disks in detached interacting
binaries, symbiotic stars and massive X-ray binaries
\citep[see e.g.][]{1984Obs...104..152L,2004MNRAS.350.1366S}.

\begin{figure}
    \centering
    \beginpgfgraphicnamed{f02}
    \begin{tikzpicture}
        \tikzstyle{post}=[->,>=stealth']
        \foreach \i in {-,+} {
            \foreach \j in {-,+} {
                \draw (\i1.2,\j1) .. controls (\i.755,\j1) and (\i.2,\j0.555) .. (0,0);
                \draw (3,0) .. controls (3.5,\j2) and (1,\j3) .. (0,\j2.95)
                .. controls (-1,\j3) and (-3.5,\j2) .. (-3,0);
            }
        \draw[rounded corners=4pt] (3,0) ..controls +(\i100:1.5cm) and +(\i40:1.5cm)
        .. (0,\i1.1) ..controls +(\i140:1.5cm) and +(\i80:1.5cm) .. (-3,0);
        }
        \draw (-1.2,1) arc(90:270:1);
        \draw (1.2,1) arc(90:-90:1);
        \draw plot[mark=x] (0,0) node [right] {$L_1$};
        \draw plot[mark=x] (3.,0) node [right] {$L_2$};
        \draw plot[mark=x] (-3.,0) node [left] {$L_3$};
        \draw plot[mark=x] (0,2.1) node [right] {$L_4$};
        \draw plot[mark=x] (0,-2.1) node [right] {$L_5$};
        \shade[ball color=white] (1.2,0) circle (.1) node [above] {$M_B$};
        \begin{scope}[xshift=-1.2cm]
            \foreach \phi in {0,30,...,360}
                \draw[post] (0,0) -- (\phi:.5);
        \end{scope}
        \shade[ball color=white] (-1.2,0) circle (.25) node {$M_A$};
        \draw (0,0) circle (4);
    \end{tikzpicture}
    \endpgfgraphicnamed
    \caption{Roche potential on the orbital plane
    showing the location of the Lagrangian
    points for a mass ratio $q=1$.}
    \label{fig:roche}
\end{figure}
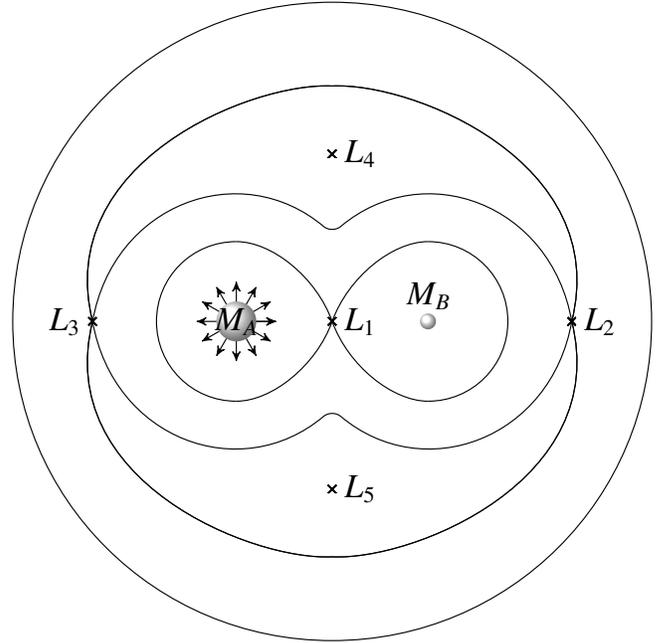

We follow the gas flow in the orbital plane of the system using 2-dimensional
hydrodynamical simulations that allows us to resolve the
large density contrasts close to the secondary.
We limit our wind accreting model to the orbital plane of the
binary system where the wind is focused forming an equatorial outflow
that has a significant density enhancement compared with the
regions further polewards \citep{2008A&A...484L...9W}.
In Fig.~\ref{fig:roche}, we show a diagram of the Roche potential in the
equatorial plane for a mass ratio $q=1$.

The basic equations of hydrodynamics describing the
evolution of the density and velocity field are:
\begin{eqnarray}
   \frac{\partial\rho}{\partial t} + \nabla \cdot (\rho \mvect{v}) & = & 0 \\
   \frac{\partial\mvect{v}}{\partial t} + (\mvect{v} \cdot \nabla ) \mvect{v} & = &
   - \frac{1}{\rho}\grad{p} - \grad{\Phi},\label{eq:NS}
\end{eqnarray}
where $\rho$ is the density in the orbital
plane, $\mvect{v}$ the velocity of the fluid, $p$ the
pressure and $\Phi$ the
gravitational potential.
Both stellar gravity potentials were approximated as
point masses.  To avoid numerical problems close to the
stars the potential is softened using the function:
\begin{equation}
\Phi = \Phi_\mathrm{A} +\Phi_\mathrm{B} = -
\frac{G M_\mathrm{A}}{\sqrt{|\mathbf{r} -
\mathbf{r}_\mathrm{A}|^2 + \epsilon^2}}
- \frac{G M_\mathrm{B}}{\sqrt{|\mathbf{r} - \mathbf{r}_\mathrm{B}|^2 + \epsilon^2}}.
\end{equation}
where 
the smoothing length $\epsilon$ was set to 0.01 length
units in our simulations with standard resolution.
The unit of length in our calculations
corresponds to the separation between the stars.
In our higher resolution simulations the number of grid zones in the accretion
region was $\sim 200$.  The surface of the compact accretor was not resolved in
our models.

\subsection{Numerical code}
\label{sec:code}

Our code is based on \flash{}
\citep{2000ApJS..131..273F}, which is a fully parallel
block-structured Adaptive Mesh Refinement (AMR)
implementation of the Piecewise Parabolic Method (PPM)
algorithm \citep{1984JCP...54..115,1984JCP...54..174} in
its original Eulerian form \footnote{The \flash{} code
is available at
\mbox{\href{http://www.flash.uchicago.edu/}{http://www.flash.uchicago.edu/}}}.  The code
has been extensively tested in various compressible flow
problems with astrophysical applications
\citep[see e.g.][]{2002ApJS..143..201C,2006MNRAS.370..529D}.

For numerical convenience we introduce dimensionless
units, where the binary separation $d$ is taken as the
unit of length. The unit of time is calculated from the
orbital angular frequency $\Omega_{\mathrm{b}}$ of the
system.
The orbital period of the system is then defined by
\begin{equation} \label{Pp}
   P_{\mathrm{b}} = 2 \, \pi. 
\end{equation}
Each simulation was run for about 10
orbital periods, when the system has reached a
quasi-equilibrium state.
The evolutionary time of the results discussed in
Section~\ref{sec:results} will be given in units of
$P_{\mathrm{b}}$.

We normalize all physical quantities for numerical
convenience.  The gravitational constant $G$ is set
to one as is done frequently in relativistic
calculations.  Eq.~\ref{eq:kepler} can be considered as a
normalization condition for the unit of mass to be the
sum of the stellar masses $M = M_{\rm A} + M_{\rm B}$.

Our implementation of the \flash{} code uses polar
coordinates and was run in both the coordinate system
corotating with the angular speed of the system and in
the inertial frame.
In addition to the usual
hydrodynamic quantities, our model includes the gravity
of both stars, and the Coriolis and centrifugal forces.
The code is based on release 2.5 of
\flash{} with customized modules for the equation of
state and gravity forces that explicitly 
conserve angular momentum transport
\citep{1998A&A...338L..37K}.
This is particularly important when large density
gradients are
present in the wind.
The Coriolis forces were treated conservatively as
described by \citet{1998A&A...338L..37K}.
A Courant number of 0.7 was used in the simulations.
We use an
equation of state for an ideal gas with ratio of
specific heats between $\gamma = 1-5/3$
to represent more realistic cases that
include radiative cooling,
and a modified isothermal
Riemann solver ported from the AMRA code
\citep{2001CPC..138..101}.

We ignore magnetic fields, radiation transfer or
explicit Navier-Stokes viscosity in our models.
\flash{} has a small numerical viscosity that has been
estimated by~\citet{2007A&A...471.1043D}.
Non-axisymmetric effects on the tidally enhanced wind may
be caused by magnetic fields.  It is likely that stellar
winds in AGB stars are primarily radial since the
extended envelope cannot have significant rotation
\citep{2002MNRAS.329..204S}.  The MHD collimation of the
wind is a very efficient mechanism and can be more
important than gravitational focusing
\citep{2000ApJ...544..336G,2001ApJ...560..928G}.
However, MHD effects are not included in our
calculations since there are presently no constraints on
the magnetic fields in symbiotic systems.  At the AGB
phase, it is unlikely that a globally ordered magnetic
field strongly affects the dynamics of the wind.

\subsection{Mass accretion onto the secondary}
\label{sec:accr}

The formation of an accretion disk can be affected by
accretion onto the secondary.  In the case of a
reflecting boundary the stream bounces off the stellar
surface and orbits the accretor. However, if no mass is
allowed to accrete onto the secondary the orbiting disk
continues to increase as the simulation continues.  We
use a mechanism that allows us to remove gas from the
vicinity of the secondary.  The accretion is accounted
for by removing some mass from the region defined by
$|\mathbf{r}-\mathbf{r}_\mathrm{B}|< r_\mathrm{acc}$
and adding it to the mass of the star to calculate the
gravitational forces.
The goal was to obtain a quasi-stable configuration and
to be able to estimate the accretion rate onto the accretor.
The size of the accretion region
$r_\mathrm{acc}$ is defined as a fraction
of the Hill radius of the mass accretor.
In most of our simulations the accretion radius is
$r_\mathrm{acc} = 0.1 R_\mathrm{H}$, where
$R_\mathrm{H}$ is the Hill radius of the secondary
\begin{equation}
    R_{\rm H} = d \left( \frac{M_{\rm B}}{3 M_{\rm{A}}}
    \right)^{1/3}.
    \label{eq:hill}
\end{equation}

The mass is removed from the disk
after each time step using the expression
\citep{2002A&A...387..550G}
\begin{equation}
   \Delta \rho = \min\left(0,f \Delta t\right) \max\left(0,\rho -
   \rho_\mathrm{av} \right),
   \label{eq:acckley}
\end{equation}
where $f$ is a constant fraction of the order unity, $\Delta t$ denotes
the timestep and
$\rho_\mathrm{av}$ is the average density in
the region
$r_\mathrm{acc}<|\mathbf{r}-\mathbf{r}_\mathrm{B}|<
2r_\mathrm{acc}$ following \citet{2008MNRAS.386..164P}.
The momentum of the accreted material is removed from
the system.

We calculate the mass accretion ratio as the ratio of
mass captured by the companion to the mass loss rate
from the primary averaged over 100 timesteps
\begin{equation}
   f= \left<
   \frac{\Mdot_\mathrm{acc}}{\Mdot_\mathrm{wind}} \right>.
   \label{eq:massacc}
\end{equation}
\citet{1996MNRAS.280.1264T}
have found accretion efficiencies around 2\%
for binaries with mass ratio $q=2$.

\subsection{Roche lobe overflow}
\label{sec:roche}

Some of the most dramatic phenomena in binary systems
occur when an AGB star fills its Roche lobe.  In this
case, a dynamically unstable mass transfer to the
companion takes place that can lead to the formation of
a common envelope and a very fast evolution of the
binary system.
We model the interaction of a tidal $L_1$ stream
with an accretion disk around a compact star
using an adiabatic equation of state and an isothermal
solver to compare the results with the wind accretion
simulations and test our numerical code.

The density of the stream flowing from the Lagrangian
inner point has the form.
\begin{equation}
   \rho = \rho_0 + \frac{\rho_\mathrm{s} - \rho_0}{\cosh((\phi/\phi_0)^\mathrm{m})}
\end{equation}
where $\rho_0$ is the initial unperturbed
density, $\rho_s$ is the density of the
stream, $\phi_0$ is the azimuthal angle of the inner
Lagrangian point and $m$ is a coefficient of order 20 that
gives the steepness of the stream.

The initial radial velocity was of the order 1\% of the
orbital velocity of the system and directed towards the
mass accreting companion. We have checked that
different initial velocities at the $L_1$ point do not affect
our results.
According to the
analytical calculations of \citet{1975ApJ...198..383L},
the tidal stream leaves the first Lagrangian point at an
incidence angle of approximately $28^{\circ}$ with
respect to the line joining the centers of the two
stars.
Since the
unit of density $\rho_0$ drops out of the equations of
motion, we may normalize it to any specified density.
It is useful to set the stream density
to a fraction of the total mass
of the system to be able to compare accretion rates to
the initial mass.

We used two different boundary conditions at the inner
edge of the grid.  The stream flow and formation of an
accretion disk depend on the interaction of the tidal
stream with the surface of the accreting star.
Therefore it is important to test different boundary
conditions at the surface of this star.  We used
reflecting boundary and allowed accretion onto the star
using the prescription defined in Section~\ref{sec:accr}
and an absorbing boundary where the radial gradients of
the velocity pressure, and density are zero at the inner
boundary.  Simulations using an absorbing boundary allow
the stream to flow into the accretor without generating
reflected waves.
We assume that the computational domain is filled with
tenuous gas with the temperature profile of the wind.
Initially, the density of the ambient material is
10 orders of magnitude smaller than the density
at the surface of the mass-losing star.

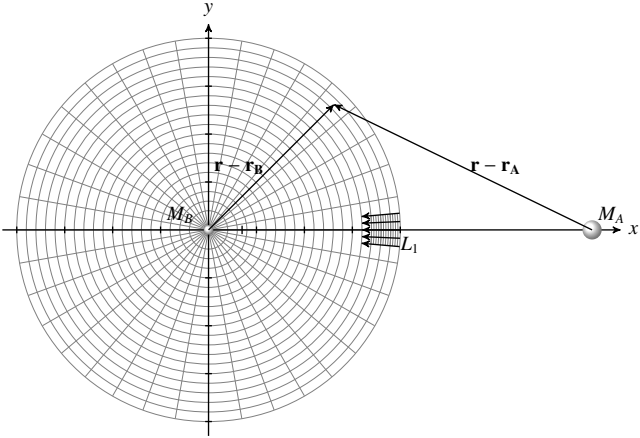
\begin{figure}
    \centering
    \beginpgfgraphicnamed{f03}
    \begin{tikzpicture}
        \tikzstyle{post}=[->,>=stealth',thick]
        \foreach \x in {.4,.6,...,4.2}
            \draw[very thin,color=gray] (0,0) circle (\x);
        \foreach \phi in {0,10,...,360}
            \draw[very thin,color=gray] (0,0) -- (\phi:4);
        \foreach \x in {3.4,3.45,...,4}
            \draw[very thin,color=gray] (\x,0) arc (0:5:\x);
        \foreach \x in {3.4,3.45,...,4}
            \draw[very thin,color=gray] (\x,0) arc (0:-5:\x);
        \draw[post] (-4.3,0) -- (8.6,0) node [right] {$x$};
        \draw[post] (0,-4.3) -- (0,4.3) node [above] {$y$};
        \foreach \x in {-4,-3,...,4} {
          \draw[thick] (\x,-2pt) -- (\x,2pt);
          \draw[thick] (-2pt,\x) -- (2pt,\x);
        }
        \draw[thick] (0.7,-2.5pt) -- (.7,2.5pt);
        \draw (4.2,0) node [below] {$L_1$};
        \shade[ball color=white] (0,0) circle (0.1);
        \shade[ball color=white] (8,0) circle (0.2);
        \draw (-.6,0) node [above] {$M_B$};
        \draw (8.4,0) node [above] {$M_A$};
        \draw[post] (0,0) -- node [left] {$\mvect{r} - \mvect{r_B}$} (45:3.7);
        \draw[post] (8,0) -- node [right] {$\mvect{r} - \mvect{r_A}$} (45:3.7);
        \foreach \y in {+,-}
        {
           \draw[post] (\y5:4) -- (\y5:3.2);
           \draw[post] (\y2.5:4) -- (\y2.5:3.2);
        }
        \draw[post] (0:4) -- (0:3.2);
    \end{tikzpicture}
    \endpgfgraphicnamed
    \caption{Grid geometry of the RLOF simulations in polar
    coordinates.
    The computational domain is centered on the accretor
    and a stream introduced over a few cells at the inner Lagrangian points.
    }
    \label{fig:gridrlof}
\end{figure}

The hydrodynamic
flow was computed in cylindrical coordinates centered on
the accreting companion and rotating in the reference
frame of the binary system.
Most of the simulations were run on a grid with 128
radial zones times 384 angular zones and 3 additional
levels of refinement.
The radial spacing in our base grid is uniform.
Fig.~\ref{fig:gridrlof}
shows a diagram of the grid geometry
and boundaries in the RLOF setup. The computational grid was centered
on the accretor and a stream is initially introduced over a few
cells at the inner Lagrangian point.
Some tests were run with twice
the linear resolution to check the convergence of the
results (see Sect.~\ref{sec:conv}).

Two different equations of state were used in our model.
One version uses an isothermal Riemann solver and the
adiabatic models use an ideal gas equation of state
with polytropic index $\gamma$ in the range $1$ to $5/3$.
The isothermal
code was used to explore the parameter space because of
its relative speed  without the need to calculate the
cooling and heating terms and evolving the energy
equation.

\subsection{Wind accretion}

In this section we describe our numerical setup
to study wind accretion in detached symbiotic binaries
where the AGB component does not fill its Roche lobe.
Many detached interacting binaries, including symbiotic
binaries and X-ray binaries show evidence for much
higher mass transfer rate than predicted by models of
spherically symmetric wind.  The enhanced accretion
rates onto the secondary can be explained by wind
focused towards the compact companion.

\subsubsection{Wind model}

\begin{figure}
   \includegraphics[width=0.5\textwidth]{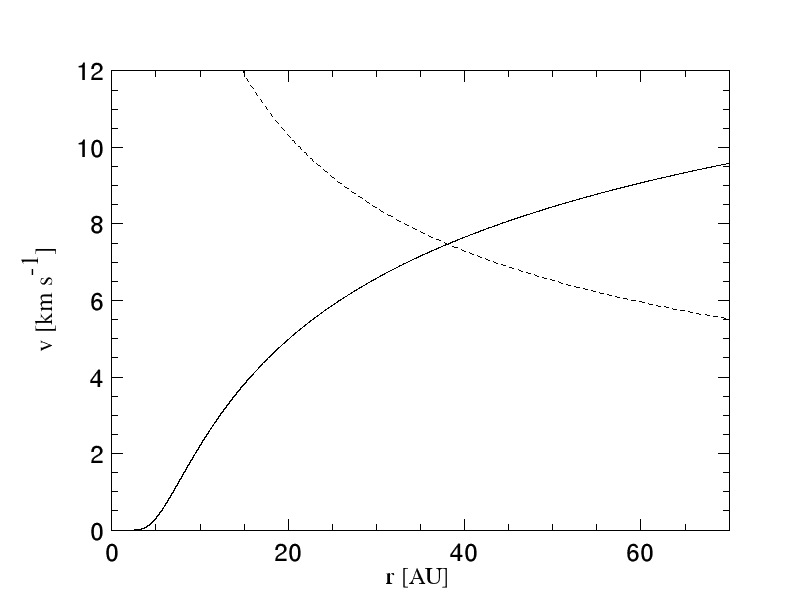}
   \caption{Radial velocity profile of the unperturbed wind
   for a temperature of $10^{3} \mathrm{K}$.
   The dotted line corresponds to the escape velocity from
   a $1.2 \Msun$ star.}
   \label{fig:wind_vel}
\end{figure}

The initial spherically symmetric wind structure is
obtained by solving the momentum equation
\begin{equation}
   v_r\frac{\ud v_r}{\ud r} + \frac{1}{\rho}\frac{\ud
   P}{\ud r} = - \frac{G M_{\rm A}}{r^2},
   \label{eq:momentum}
\end{equation}
where $v_r$ is the radial component of the wind velocity and
$M_{\rm A}$ the mass of the central star. The solution for an
isothermal wind is shown in Fig.~\ref{fig:wind_vel}.  As
the wind propagates across the grid it is accelerated
by thermal pressure and its velocity increases.
It is possible that AGB winds are not
completely spherically symmetric due to internal
processes in the star or the influence of the secondary
\citep{1996A&A...313..605D}.

\begin{figure}
    \resizebox{\hsize}{!}{\includegraphics{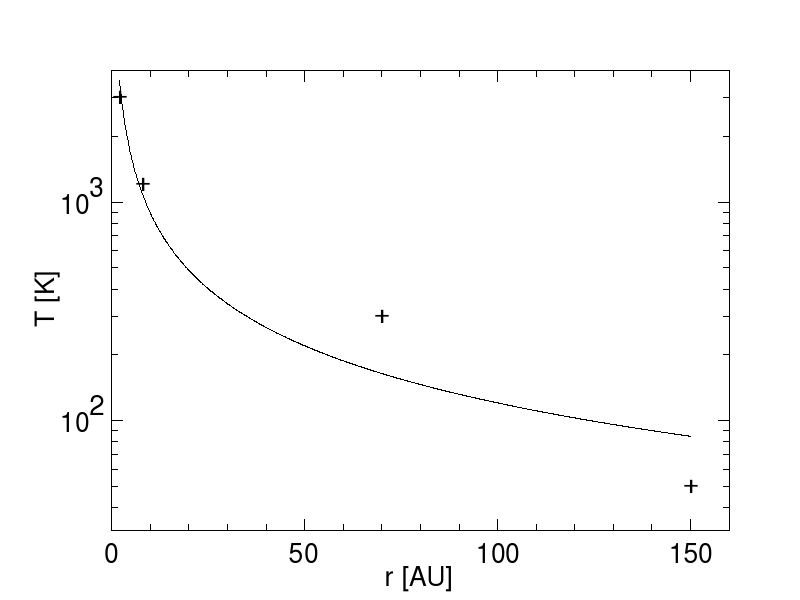}}
   \caption{Temperature as a function of distance from the
   AGB star in logarithmic scale used in our calculations
   for different wind acceleration radii. The temperature
   is   $\sim 10^3 \mathrm{K}$ at $10 \mathrm{AU}$
   where dust is formed. The crosses give dust temperature
   estimates from IR observations of the symbiotic
   system Mira~AB \citep{2001ApJ...556L..47M}.
   }
   \label{fig:wind_temp}
\end{figure}

\begin{figure}
    \resizebox{\hsize}{!}{\includegraphics{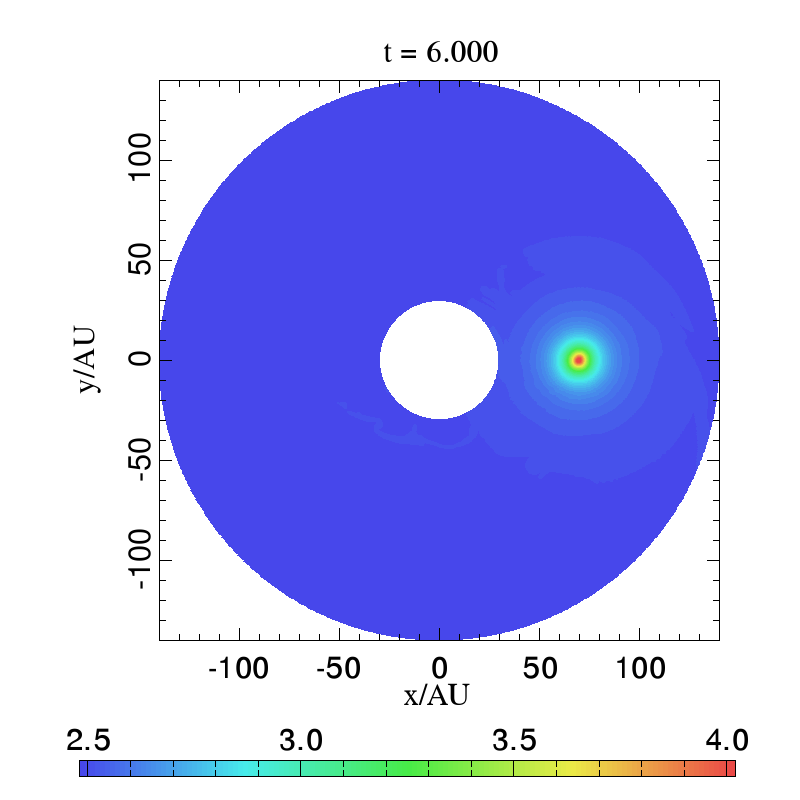}}
   \caption{Temperature distribution in logarithmic
   scale after 6 orbits
   for a binary system with separation of $70 \mathrm{AU}$.
   }
   \label{fig:temp}
\end{figure}

In Fig.~\ref{fig:wind_temp} we plot the temperature
profile used in our wind model.  During the AGB phase
it is likely that the winds are primarily radial since
the massive stellar atmosphere cannot have significant
rotation.  Our temperature distribution has a radial
profile $\propto r^{-0.5}$ and is similar to
the one given by the AGB models by
\citet{1997MNRAS.287..799I}.
Fig.\ref{fig:temp} shows the temperature distribution 
after 6 orbits in our locally isothermal simulations.

The binary systems considered here consist of a
$1.2\,M_{\sun}$ AGB star and a $0.6\,M_{\sun}$
companion.  The orbital separation $d$ ranges between
$10\,\mathrm{AU}$ and $70\,\mathrm{AU}$.
The initial
temperature of the wind is between $3000\,\mathrm{K}$
and $10000\,\mathrm{K}$ and the velocity of the wind on
the surface of the AGB star is $2-15 \kms$.  Both
temperature and the radial velocity of the wind are
constant at the dust formation radius for a given wind
model.

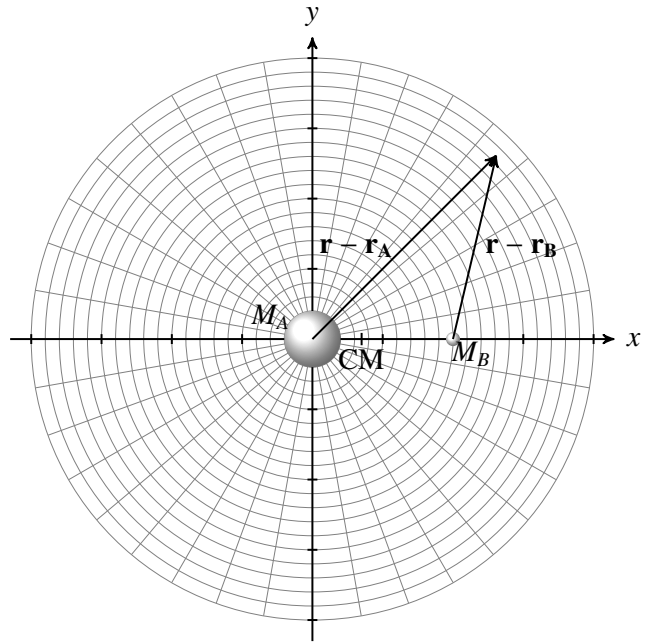
\begin{figure}
    \centering
    \beginpgfgraphicnamed{f06}
    \begin{tikzpicture}
        \tikzstyle{post}=[->,>=stealth',thick]
        \foreach \x in {.4,.6,...,4.2}
            \draw[very thin,color=gray] (0,0) circle (\x);
        \foreach \phi in {0,10,...,360}
            \draw[very thin,color=gray] (0,0) -- (\phi:4);
        \draw[post] (-4.3,0) -- (4.3,0) node [right] {$x$};
        \draw[post] (0,-4.3) -- (0,4.3) node [above] {$y$};
        \foreach \x in {-4,-3,...,4} {
          \draw[thick] (\x,-2pt) -- (\x,2pt);
          \draw[thick] (-2pt,\x) -- (2pt,\x);
        }
        \draw[thick] (0.7,-2.5pt) -- (.7,2.5pt);
        \draw (.7,0) node [below] {CM};
        \shade[ball color=white] (0,0) circle (0.4);
        \shade[ball color=white] (2,0) circle (0.1);
        \draw (-.6,0) node [above] {$M_A$};
        \draw (2.25,-pi/15) node {$M_B$};
        \draw[post] (0,0) -- node [left] {$\mvect{r} - \mvect{r_A}$} (45:3.7);
        \draw[post] (2,0) -- node [right] {$\mvect{r} - \mvect{r_B}$} (45:3.7);
    \end{tikzpicture}
    \endpgfgraphicnamed
    \caption{Schematic representation of the grid geometry in polar
    coordinates.  The physical quantities are defined
    in the center of the cells.
    The system is centered on the primary and
    rotating in clockwise direction.}
    \label{fig:grid}
\end{figure}

The computational domain was centered on the primary and
the secondary was fixed on a circular orbit
neglecting the hydrodynamic effects.
Our grid extended out to twice the binary
separation in our models which is enough to avoid
mass loss from the disk around secondary.
We used a grid with a fixed
uniform size in the radial and azimuthal directions.
The grid cells were sized to give square cells at the
location of the secondary's orbit.  To improve the
resolution close to the secondary additional levels of
refinement were used to speed up the simulation.  We
introduce the mass of the secondary over 1 orbital
period to avoid the formation of strong shocks.  The
adaptive mesh refinement allows us to resolve a broad
range of density contrasts close to the secondary.
A diagram of the grid geometry
is shown in Fig.~\ref{fig:grid}.

We assume that the wind is driven by radiation pressure
and is composed of dust and a photoionized ideal
monoatomic gas which are perfectly coupled.  The wind
acceleration mechanism is not considered in our model.
The wind is accelerated at the dust condensation radius
according to the Eq.~\ref{eq:momentum}
where the full gravity of the AGB star is included.
We include gas forces due to pressure and gravity
from the AGB and the compact companion but the
self-gravity of the gas is neglected.
In our adiabatic runs, we assume that the gravity of the
red giant is balanced by radiation pressure on the dust,
so that it does not appear explicitly in our equations.
Dust grains form in the AGB envelope,
where pulsational shocks bring material from the
atmosphere, and deposit it in the region of
$\sim 2$ photospheric radii.
In this case the only
force acting on the wind from the AGB star is the gas
pressure, whereas in a binary the wind is additionally
subject to the gravity of the companion.  Such an
approximation leads to terminal velocities of the wind
ranging between $10-40 \kms$, comparable to those obtained by
\cite{2000A&A...361..641W} for winds in which radiation
pressure nearly balances the gravity of the wind source
\citep[see e.g.][]{2002A&A...385..205G}.  The effects of cooling are
accounted for by a nearly isothermal equation of state
in some of our runs, while in the adiabatic runs
the polytropic index $\gamma$ is in the range $1-5/3$.

We neglect radiation pressure from the secondary.  In
the case of a white dwarf companion, for typical dust
grain sizes the effect of radiation pressure on grains
is negligible compared with gravity focusing.
Explicit radiative heating or cooling terms are also
neglected in our wind model.

\subsubsection{Equation of state}

The pressure is obtained by
\begin{equation}
   p = \frac{\mathcal{R} \rho T}{\mu},
\end{equation}
where $\mathcal{R}$ is the gas constant, $T$ is the
temperature and $\mu$ is the mean atomic weight.

The temperature is approximately constant at the
secondary's position. Because of the gravitational field
of the accretor, the disk aspect ratio calculated with
respect to the $h_\mathrm{B} = H/|\mathbf{r} -
\mathbf{r}_\mathrm{B}|$ can decrease considerably.  This
influences the flow in vicinity of the secondary and
increases the amount of gas accumulated inside the Hill
sphere. On the other hand we should expect a temperature
increase in the vicinity of the secondary.
For of a permanent disk
$h_\mathrm{B}$ can increase due to the compression of
the gas flowing into the vicinity of the accretor.  We
use the following expression to calculate the sound speed
\begin{equation}
\label{cs_binary}
c_\mathrm{s}={{h_\mathrm{A} r_\mathrm{A} h_\mathrm{B} r_\mathrm{B}} \over {\left((h_\mathrm{A} r_\mathrm{A})^n+(h_\mathrm{B} r_\mathrm{B})^n\right)^{1/n}}}{\sqrt{\frac{G M_\mathrm{A}}{r_\mathrm{A}^3}
+ \frac{G M_\mathrm{B}}{r_\mathrm{B}^3}}},
\end{equation}
where $r_\mathrm{A}= |\mathbf{r} -
\mathbf{r}_\mathrm{A}|$ is the distance from the
mass-losing primary, $r_\mathrm{B}= |\mathbf{r} -
\mathbf{r}_\mathrm{B}|$ is the distance from the
secondary, and a parameter $n=2$ is used in our
simulations.

A disk with constant aspect ratio $h_\mathrm{B}$ can be
formed around the mass gainer with
a Keplerian angular velocity
\begin{equation}
\Omega_\mathrm{B} = \sqrt{G M_\mathrm{B} \over r_\mathrm{B}^3}.
\end{equation}

\subsubsection{Model parameters}

The main parameters of the models are the wind
temperature that determines its terminal velocity, the
orbital parameters (binary separation and orbital period
for a circular orbit), mass loss rate from the AGB star
and the mass ratio defined by
\begin{equation}
    q = \frac{M_\mathrm{A}}{M_\mathrm{B}},
   \label{eq:massratio}
\end{equation}
where $M_{\rm A}$ is the mass of the primary mass-losing star
and $M_{\rm B}$ is the mass of the wind-accreting compact
star.

Here we describe the model parameters of the runs
discussed in Section~\ref{sec:results}.  The indexing of
the models described below increases with increasing
dust condensation radius to facilitate reference to models.
The average wind terminal velocity from an AGB star is
around 15 \kms, within a range of 10 \kms.  In the
following, we produce models having a terminal velocity
for a single star between 10 and 25 \kms.  The binary
separations $d$ vary between 10 and 70 AU.
Our base resolution with 3 levels of refinement 
corresponds to a cell size of $\sim 0.004$ close
to the accretor.
Each simulation with our standard resolution requires
about 30 hours of CPU time on Pentium 4 processors.

The inner Lagrangian point in model
M-4 is at 40 AU from the mass-lossing star, which is
close to the inner boundary of the domain.  In this case
we expect to see some of the
effects of RLOF in addition to that of wind accretion.

\begin{table}
   \caption[]{Model parameters of a long-period and
   close symbiotic binaries.}
      \label{tbl:model}
      \centering
      \begin{tabular}{lcc}
            \hline \hline
            Parameter &  long-period & short-period\\
            \hline
            Mass of primary $M_\mathrm{A}$ (g) &
            $2.387\times10^{33}$  &
            $2.387\times10^{33}$ \\
            Mass of secondary $M_\mathrm{B}$ (g) &
            $1.193\times10^{33}$  &
            $1.193\times10^{33}$ \\
            Radius of primary $R_\mathrm{A}$ (cm) &
            $2.394\times10^{14}$ &
            $2.394\times10^{14}$ \\
            Radius of secondary $R_\mathrm{B}$ (cm) &
            $1.496\times10^{14}$ &
            $1.496\times10^{14}$ \\
            Wind temperature $T_\mathrm{wind}$ (K) &
            1000 & 1200\\
            Binary separation $d$ (cm) &
            $1.047\times10^{15}$ & $2.394\times10^{14}$\\
            Period $P_{\rm b}$ (yr) & 436.5 & 40.9 \\
            \hline
      \end{tabular}
\end{table}

The parameters in these simulations are closely related
to two symbiotic systems Mira~AB and R~Aqr (see
Table~\ref{tbl:model}).
The wide symbiotic system Mira~AB (\object{$o$
Ceti}) consists of a pulsating AGB primary (Mira) and a
compact secondary at an \emph{Hipparcus} distance of $\sim 128
\mathrm{pc}$.
Mira stars are long-period variable stars which evolve along
the AGB and present large stellar pulsations.  The
primary's radius is found to be $\sim 270 \Rsun$.
The nature of the
companion is uncertain and has been suggested to be a
white dwarf or a late-type, magnetically active
main-sequence dwarf.
HST observations of Mira~AB have shown that it is a strongly interacting
system with mass transfer from the primary towards the compact companion.
Its optical through UV spectrum
appears to be dominated by emission from an accretion
disk that presumably is accumulated from Mira A's
massive wind
\citep{1985ApJ...297..275R,1997ApJ...482L.175K}.  Recent
observations of symbiotic binaries in wavebands ranging
from the IR to the X-rays suggest that wind focusing can
form a stream flow even in wide systems
\citep{1997ApJ...482L.175K,2005ApJ...623L.137K}.
The system has large variations in
luminosity 
with well-defined pulsation periods.  X-rays from the
Mira AB system arise either from magnetospheric
accretion of wind material from Mira A onto Mira B, or
from coronal activity associated with Mira B itself, as
a consequence of accretion-driven spin-up.

\object{R~Aqr} is a close symbiotic binary system
(at a distance of $\sim 200 \mathrm{pc}$)
composed of a mass-losing Mira variable and a white dwarf
surrounded by an extended nebula
\citep{1985A&A...148..274S}.  The orbital period is not
well determined with values in the literature ranging
between 20 and 44 yr
\citep{1981IBVS.1961....1W,1989AJ.....98.1820H,2007AJ....134.2113M}.
The companion is surrounded by an accretion disk which
gives rise to a jet with bipolar outflow.  The primary
almost fills  its Roche lobe during periastron, although
mass flow towards the companion is increased
due to wind focusing during the rest of the orbital motion
\citep{2008arXiv0804.4139G}.

\section{Results and Discussion}
\label{sec:results}

The wind parameters and mass ratio of our models of wind
accreting symbiotic binaries are given in
Table~\ref{tbl:mira-models}.
We will consider mainly the simulations with a locally isothermal
equation of state in the following discussion.
The adiabatic runs will be
considered in Sec.~\ref{sec:resultsgamma}.

\subsection{Roche lobe overflow}

\begin{figure*}
   \begin{tabular}{cc}
      \includegraphics[width=0.5\textwidth]{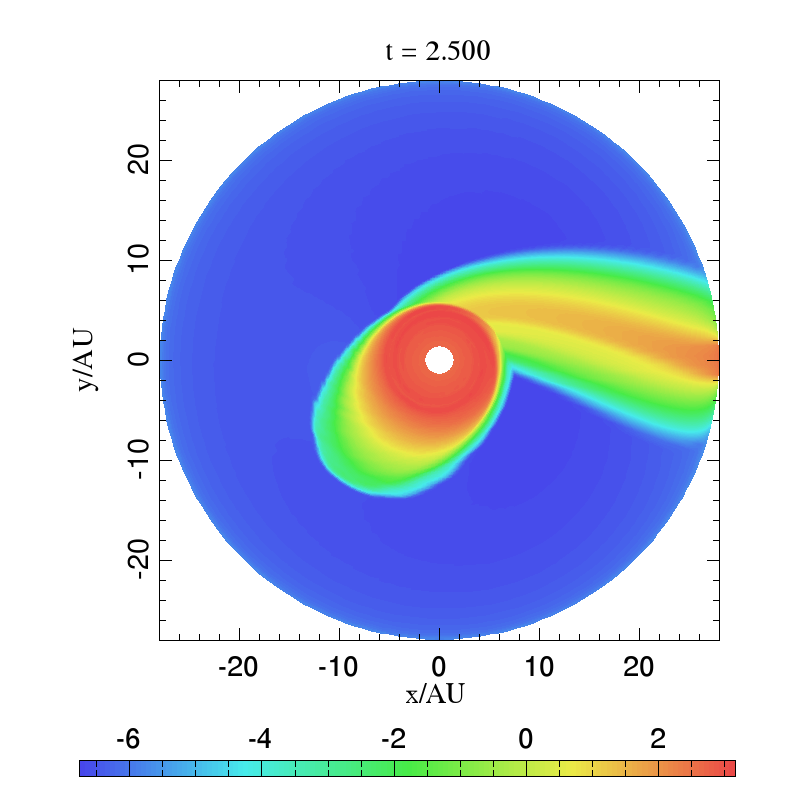} &
      \includegraphics[width=0.5\textwidth]{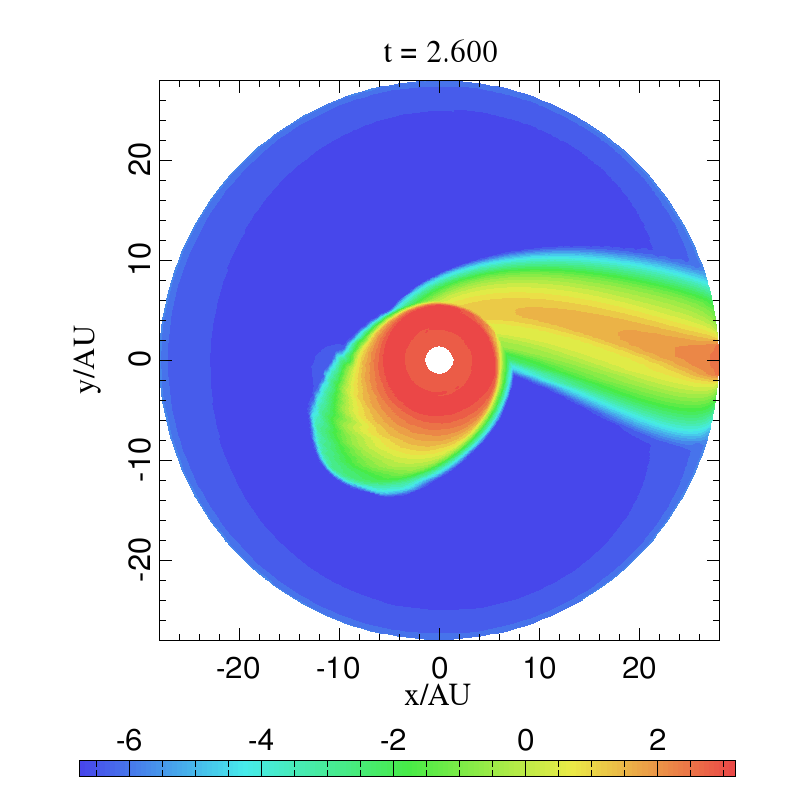} \\
   \end{tabular}
   \caption{Density contours in the orbital plane
   of an RLOF simulation   for an adiabatic run with $\gamma=1.1$ (left) and a
   locally isothermal run (right).
   The system has a separation of $70 \mathrm{AU}$
   and a mass ratio $q=2$.}
   \label{fig:Roche}
\end{figure*}

First, we study the accretion in the RLOF case to check
our code and compare the results with the wind accreting
cases described below in Sec.~\ref{sec:resultswind}.
Also, in some symbiotics RLOF is likely to occur during
periastron if their orbits are highly
elliptical, for example in R~Aqr where the
stellar separation becomes 8 AU \citep{2008arXiv0804.4139G}.
The gas from the primary star is assumed to flow through
the inner equilibrium point with an initial velocity
between $0.01-0.1$ of the orbital velocity.  We run our
simulations for 5 orbital periods, when the system has
reached a quasi-steady state.

\begin{figure}
    \includegraphics[width=0.5\textwidth]{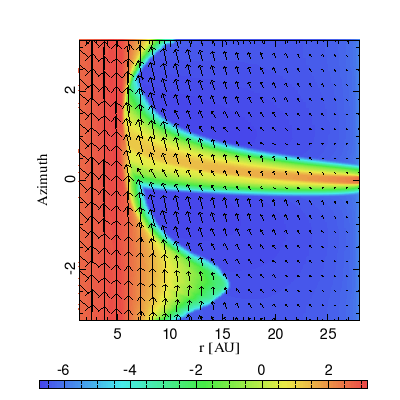}
    \caption{Density contours in the orbital plane
    in polar coordinates for the RLOF simulations after 1 orbit.
    Velocity vectors are shown in the inertial frame .
    }
    \label{fig:rlofvectors}
\end{figure}

Our simulations show that most of the flow is
efficiently accreted onto the star as it turns around
and forms a stable disk with counterclockwise rotation
and elliptic streamlines.  The gas is accelerated by the
companion's gravity and Coriolis force and becomes
supersonic. The $L_1$ stream cannot hit the companion
star due to the Coriolis force and penetrates into the
disk forming a hot spot that heats up the outer disk in
our adiabatic runs with $\gamma = 5/3$.

In Fig.~\ref{fig:Roche} we show the density contours for
a system with separation of $70 \mathrm{AU}$ and $L_1$
Lagrangian point located at $\sim 30 \mathrm{AU}$ from
the accretor using two different equations of state.
Mass is removed inside 0.5 Hill radius in our
simulations to achieve a quasi-stable state after
several orbital periods.  The angle of deflection of the
stream roughly agrees with the analytical estimates of
\citet{1975ApJ...198..383L}.
Fig.~\ref{fig:rlofvectors} shows the velocity vectors in the
inertial frame for one of the RLOF runs.

The flow patterns are non-axisymmetric due to the shock
waves that form when gas rotating in the disk collide
with the stream. This kind of patterns have been
observed using Doppler tomography in the disk
around IP
Pegasi~\citep{1997MNRAS.290L..28S,2000MNRAS.316..906M}.
The angular momentum transfer in the
disk is more efficient due to the presence of shocks and
accretion may take place without turbulent viscosity.
There is some mass loss in our grid in the case of
reflective boundaries at the accretor's surface.

Since cooling effects due to radiation are not included
in our equation of state we use a locally isothermal
model where the temperature decreases as $r^{-1}$
from the center of the secondary.
In this case the stream shocks are stronger
than in  the adiabatic runs
(right panel in Fig.~\ref{fig:Roche}).

\subsection{Wind accretion}
\label{sec:resultswind}

\begin{figure*}
   \begin{tabular}{cccc}
      \includegraphics[width=0.24\textwidth]{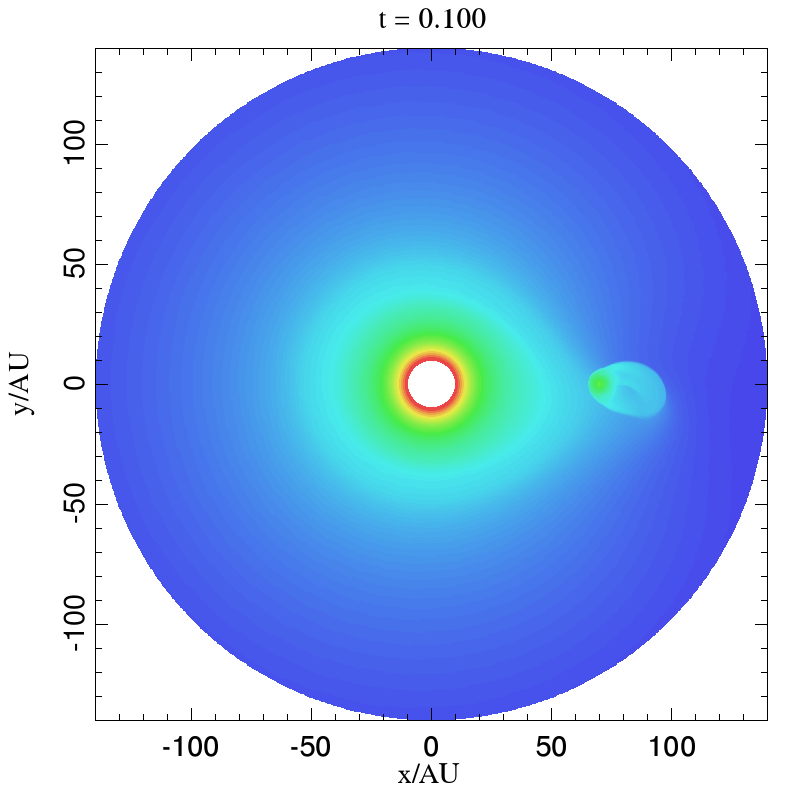} &
      \includegraphics[width=0.24\textwidth]{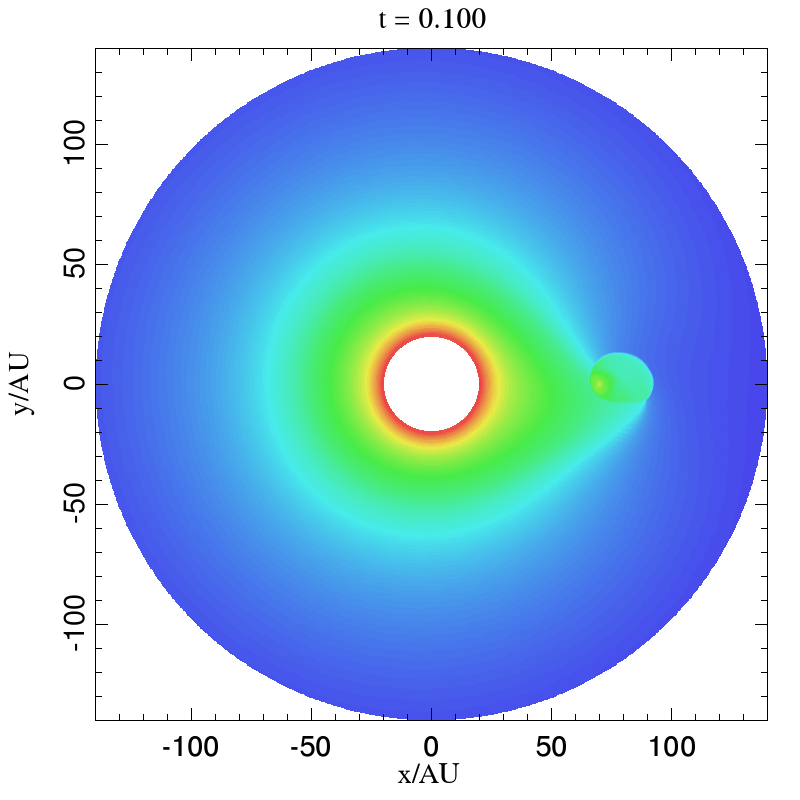} &
      \includegraphics[width=0.24\textwidth]{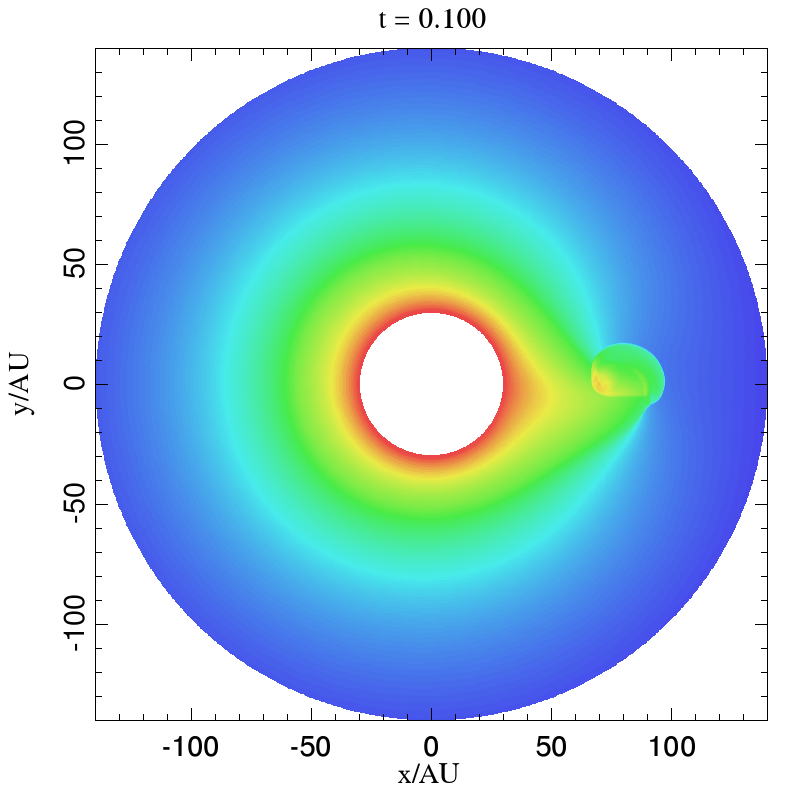} &
      \includegraphics[width=0.24\textwidth]{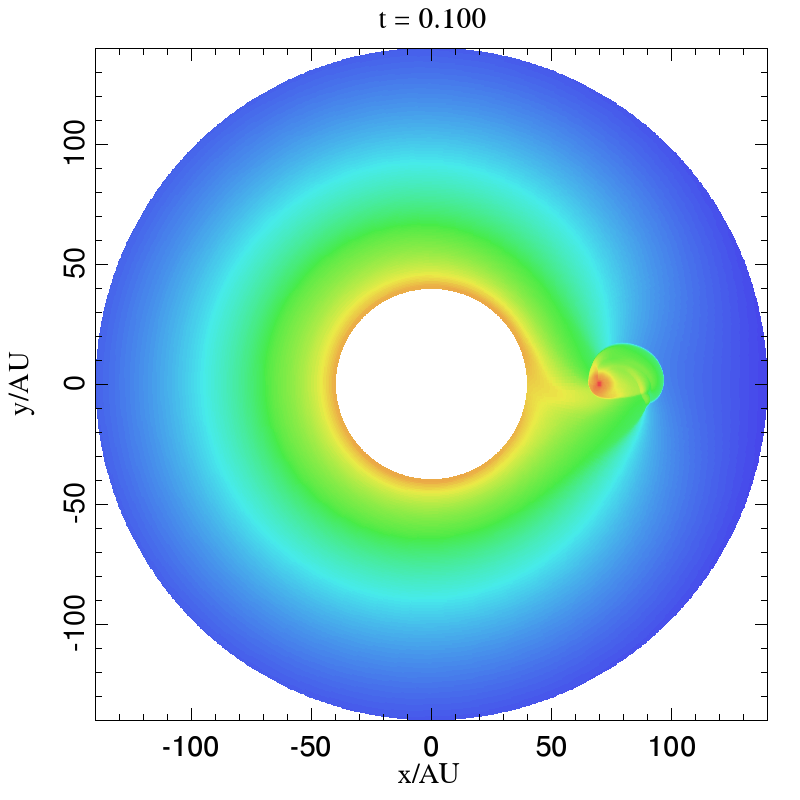}
\\
      \includegraphics[width=0.24\textwidth]{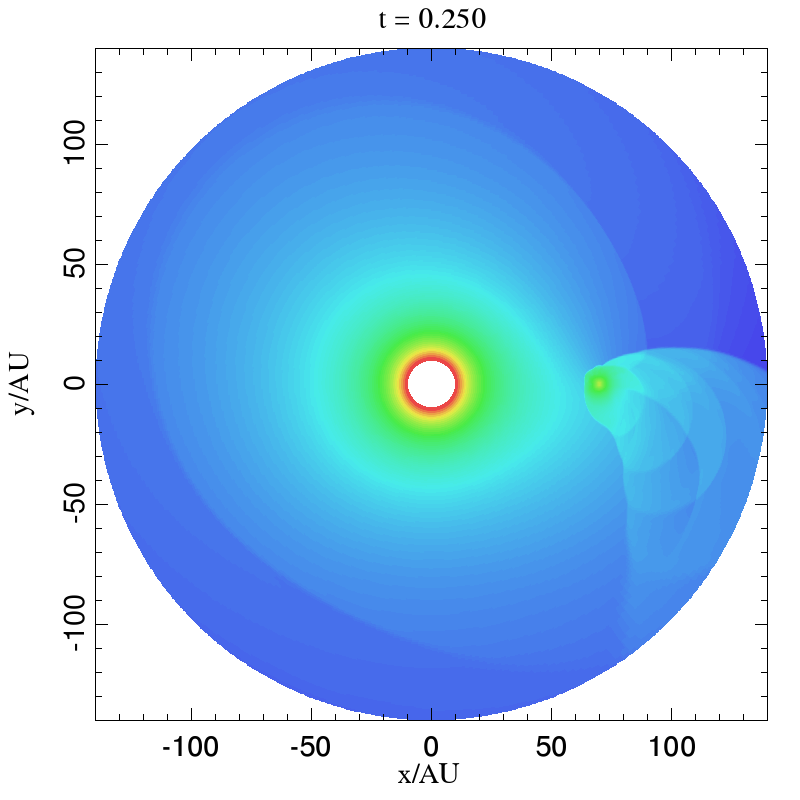} &
      \includegraphics[width=0.24\textwidth]{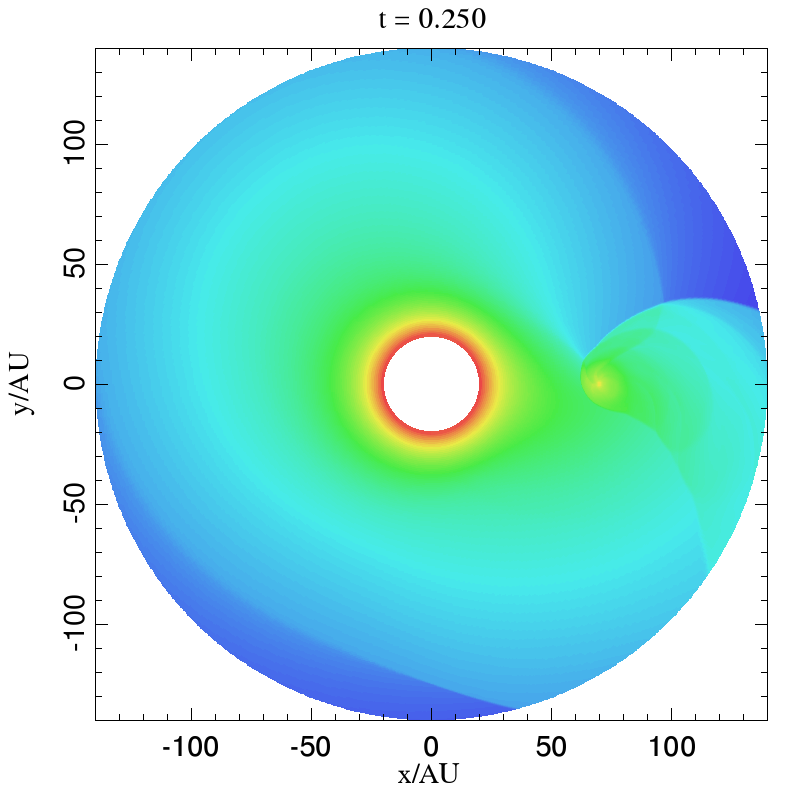} &
      \includegraphics[width=0.24\textwidth]{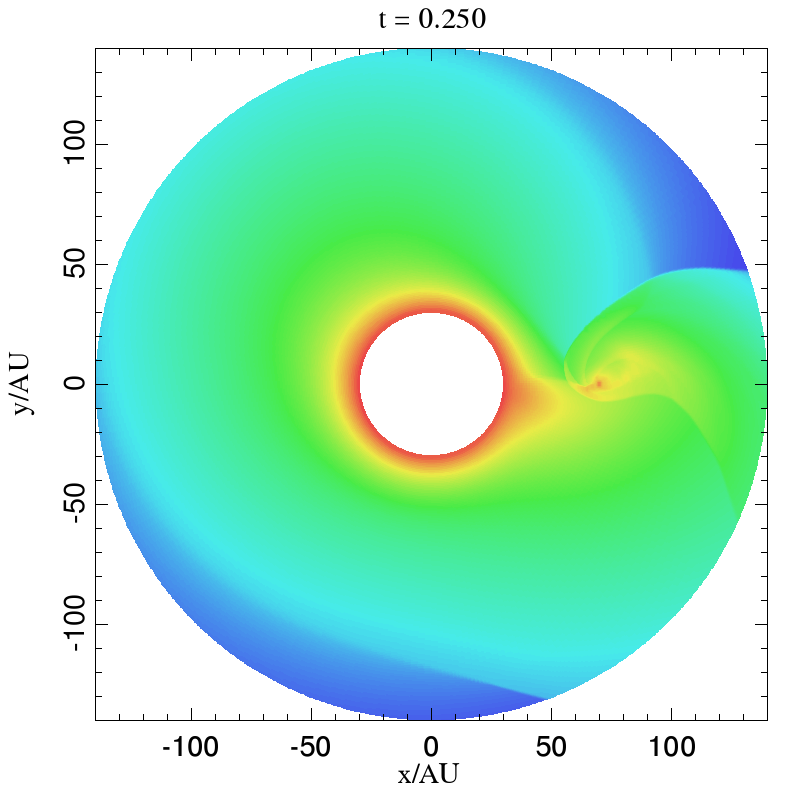} &
      \includegraphics[width=0.24\textwidth]{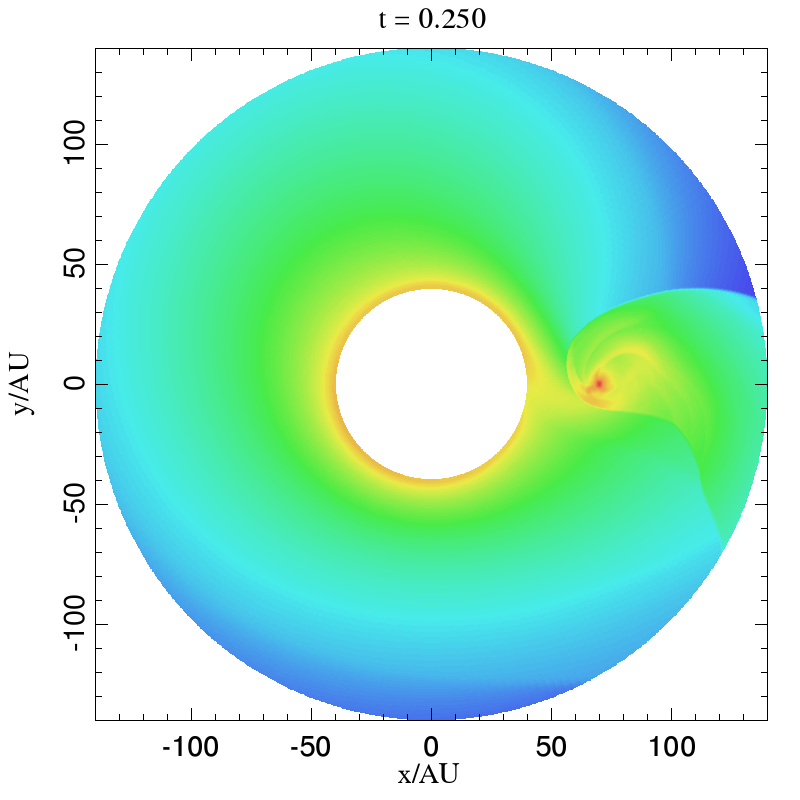}
\\
      \includegraphics[width=0.24\textwidth]{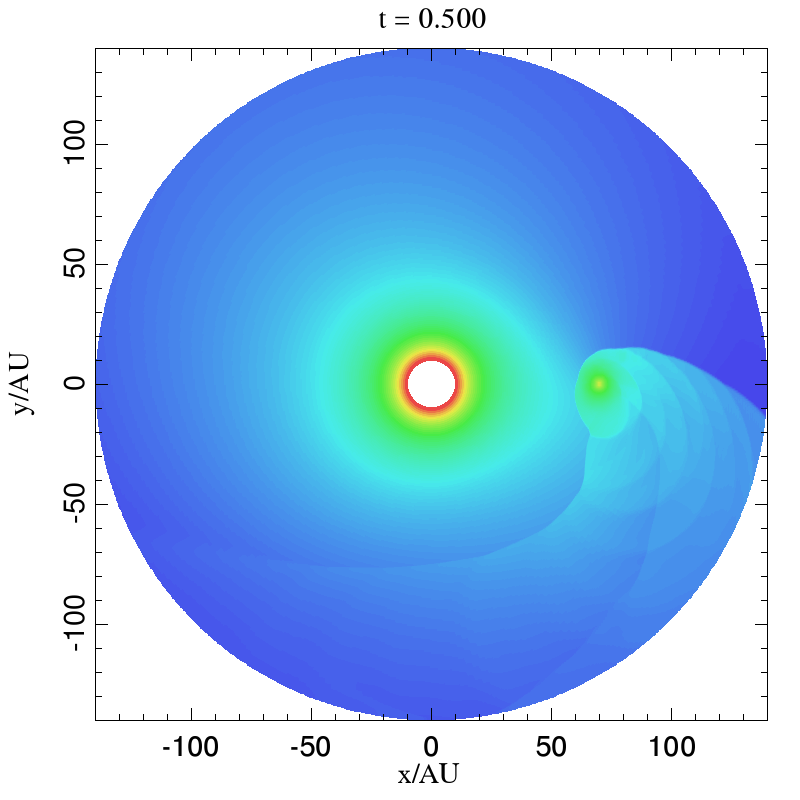} &
      \includegraphics[width=0.24\textwidth]{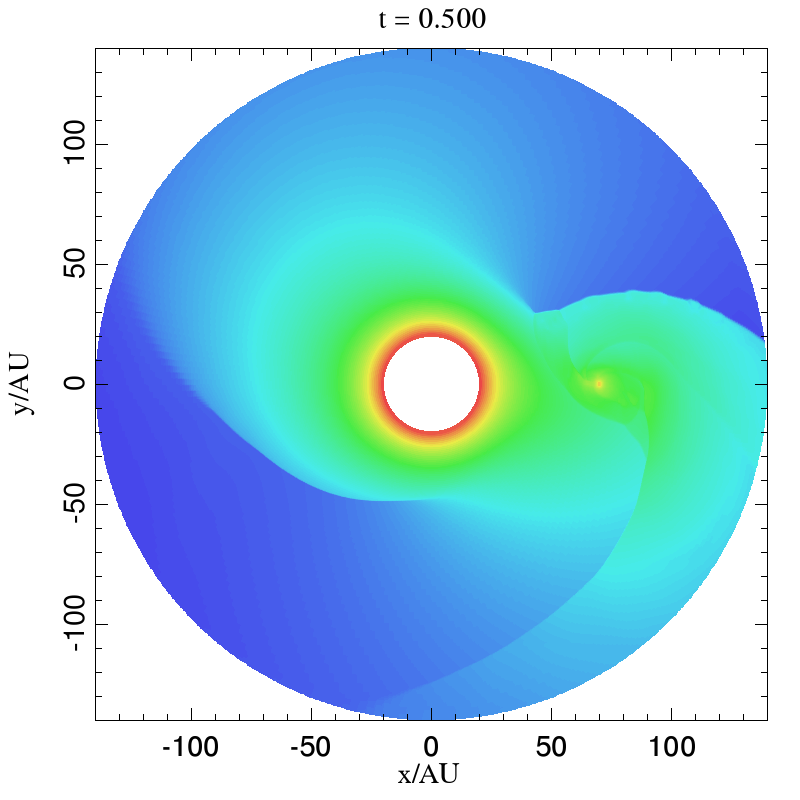} &
      \includegraphics[width=0.24\textwidth]{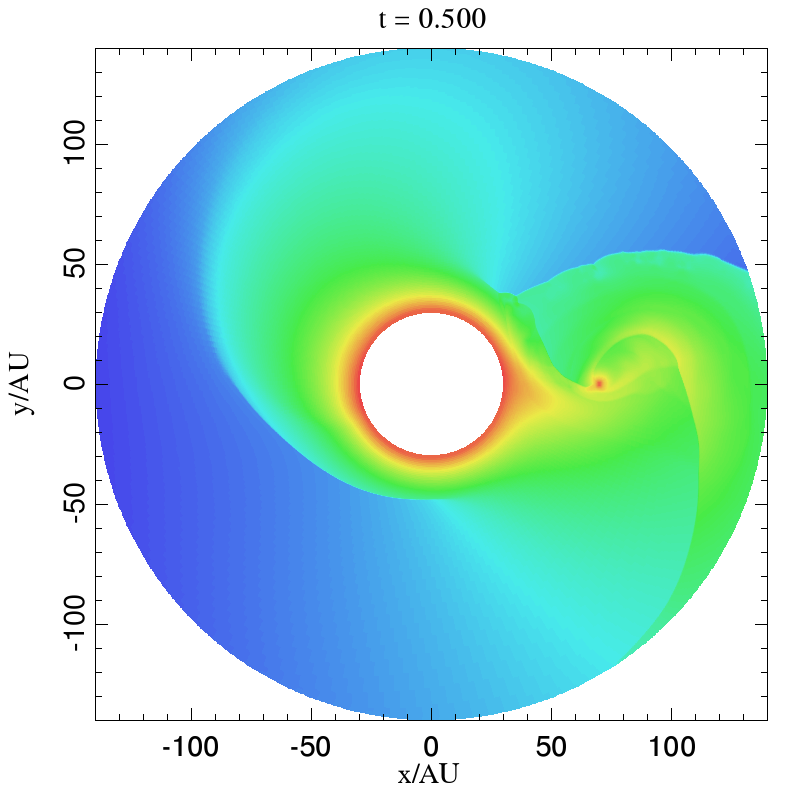} &
      \includegraphics[width=0.24\textwidth]{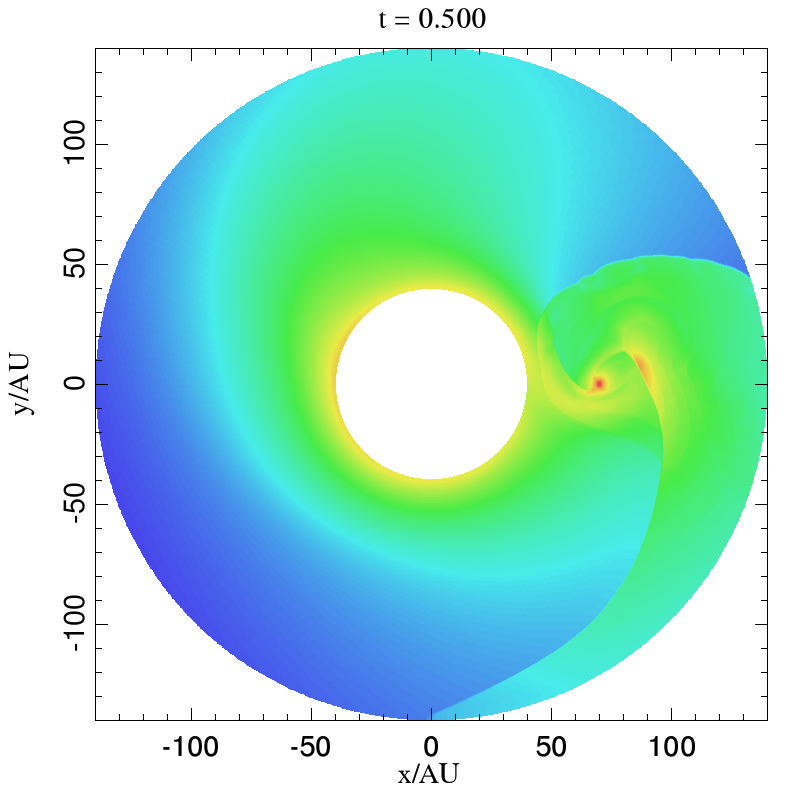}
\\
      \includegraphics[width=0.24\textwidth]{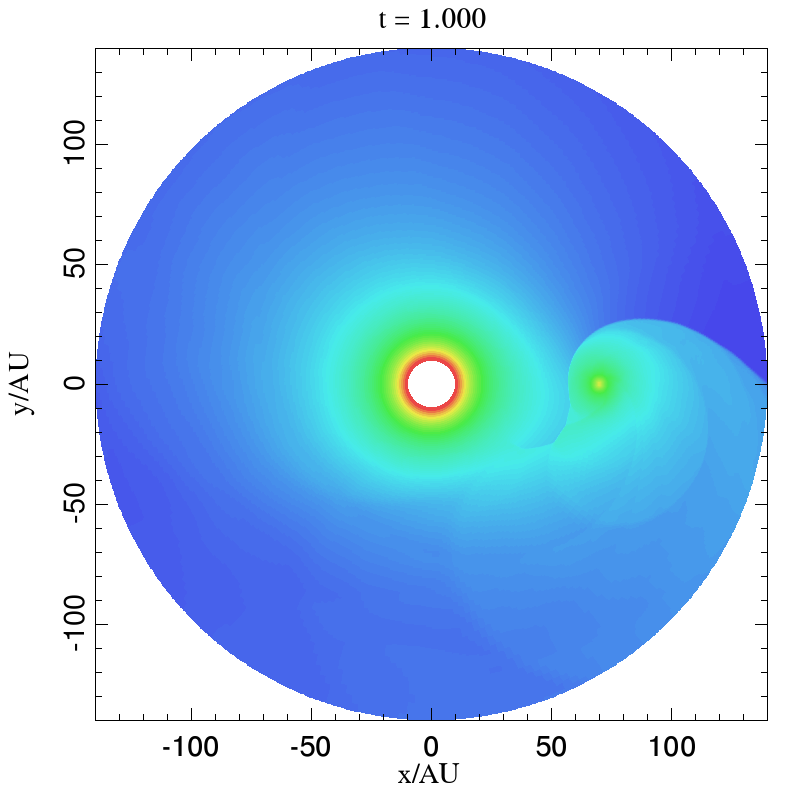} &
      \includegraphics[width=0.24\textwidth]{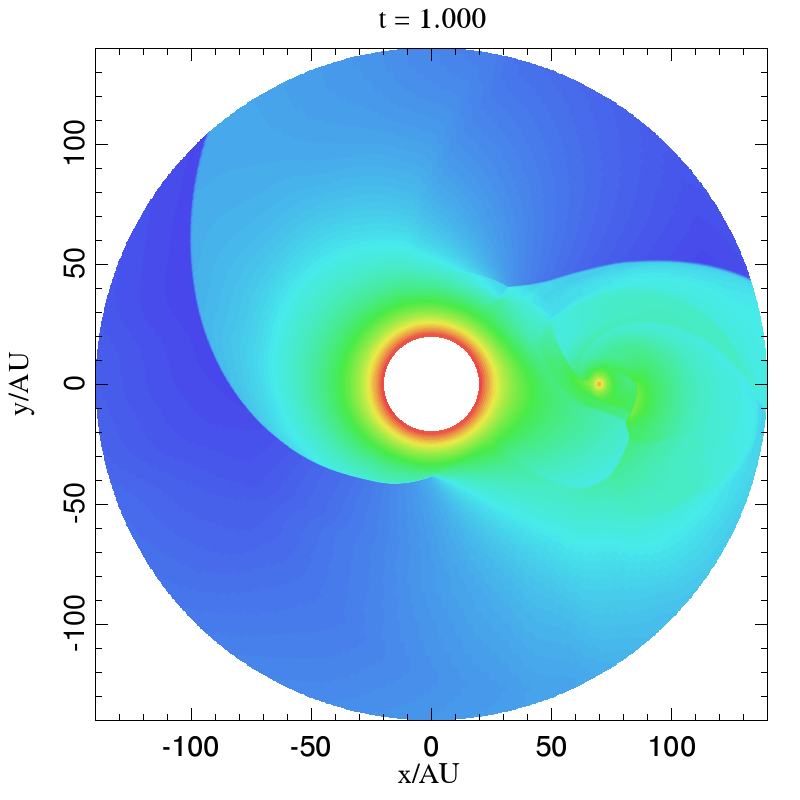} &
      \includegraphics[width=0.24\textwidth]{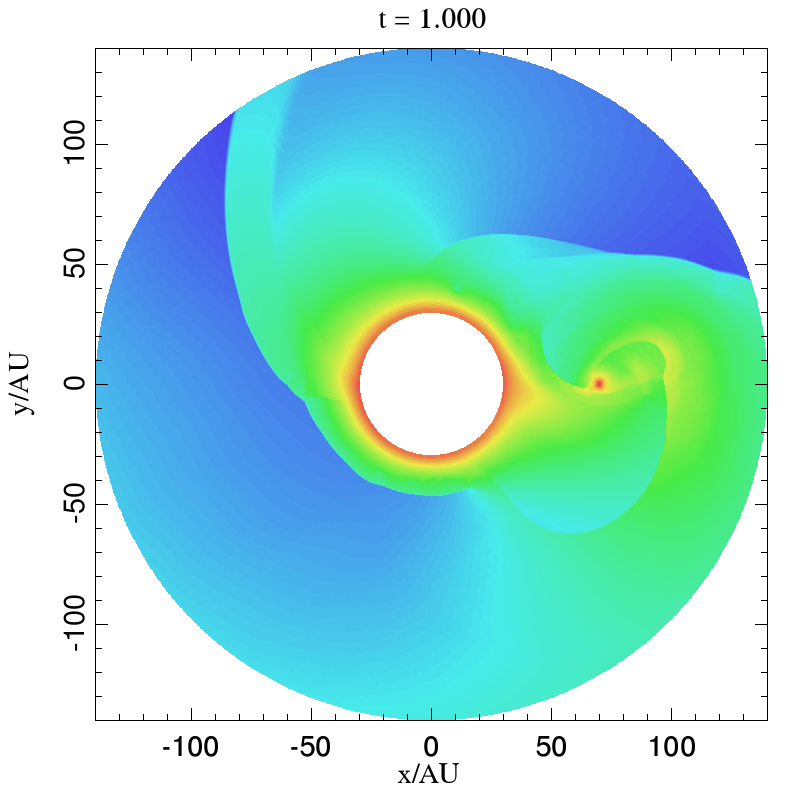} &
      \includegraphics[width=0.24\textwidth]{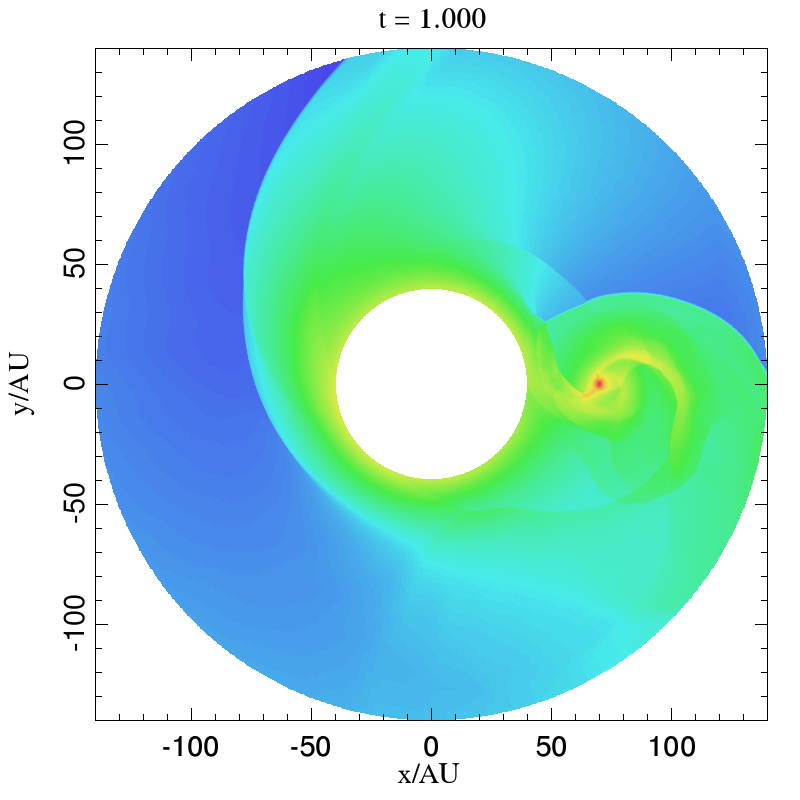}
   \end{tabular}
   \caption{Density contours
   in logarithmic scale
   for a separation of $70 \mathrm{AU}$
   at several times.
   From left to right
   the inner boundary where the wind is accelerated
   is $10, 20, 30$ and $40 \mathrm{AU}$.
   The model parameters are
   presented in Table~\ref{tbl:mira-models}.
   In the outermost right column the envelope of the primary
   is close to the Hill surface.
   Three snapshots are shown for times $t = 0.25, 0.5$ and $1$
   orbital periods.
   The temperature of the wind decreases roughly as $r^{-1}$ where $r$
   is the distance from the primary.
     }
   \label{fig:miratime}
\end{figure*}

Our results indicate that wind accretion in
symbiotic binaries can result in RLOF-like accretion
via a stream flow.
Stable accretion disks can form for
winds with mass loss ranging between
$10^{-9}-10^{-6} \Msun \yr$
\citep{2000ApJ...545..945R}
and distances between $16-70
\mathrm{AU}$.

\begin{figure}
   \includegraphics[width=0.4\textwidth]{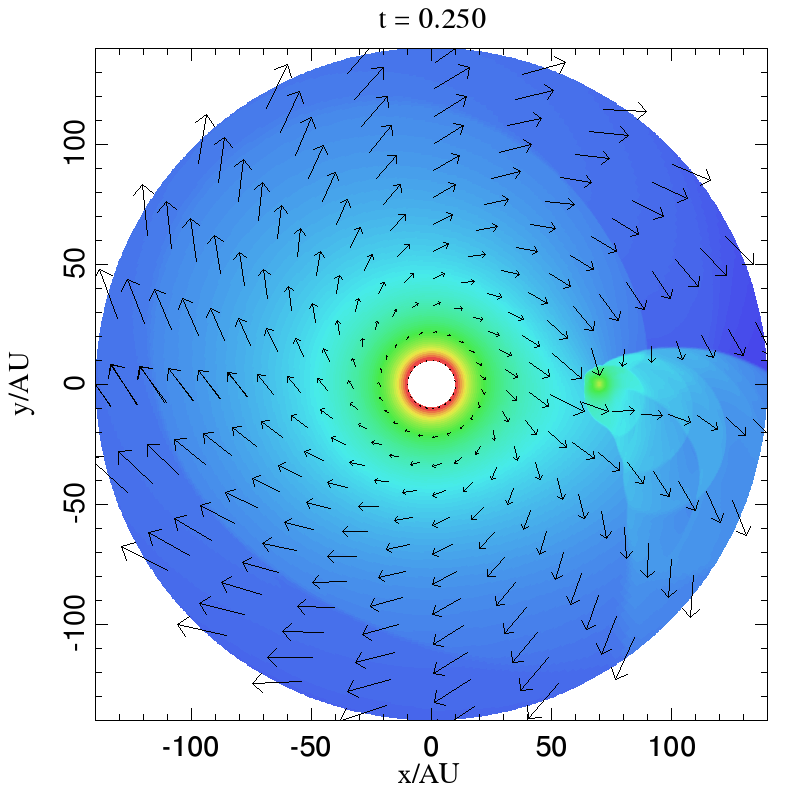}
   \includegraphics[width=0.4\textwidth]{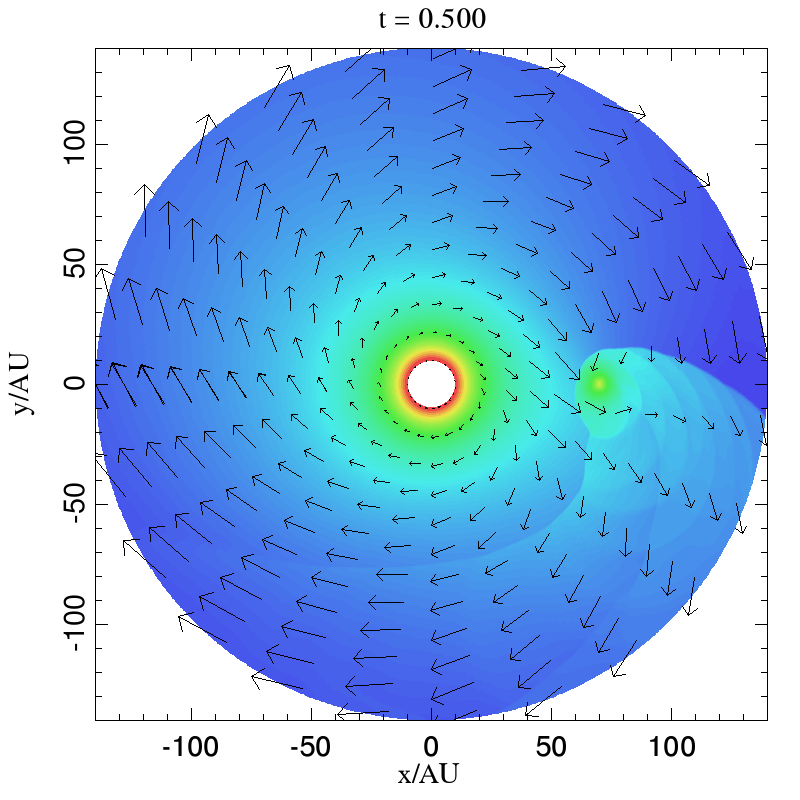}
   \includegraphics[width=0.4\textwidth]{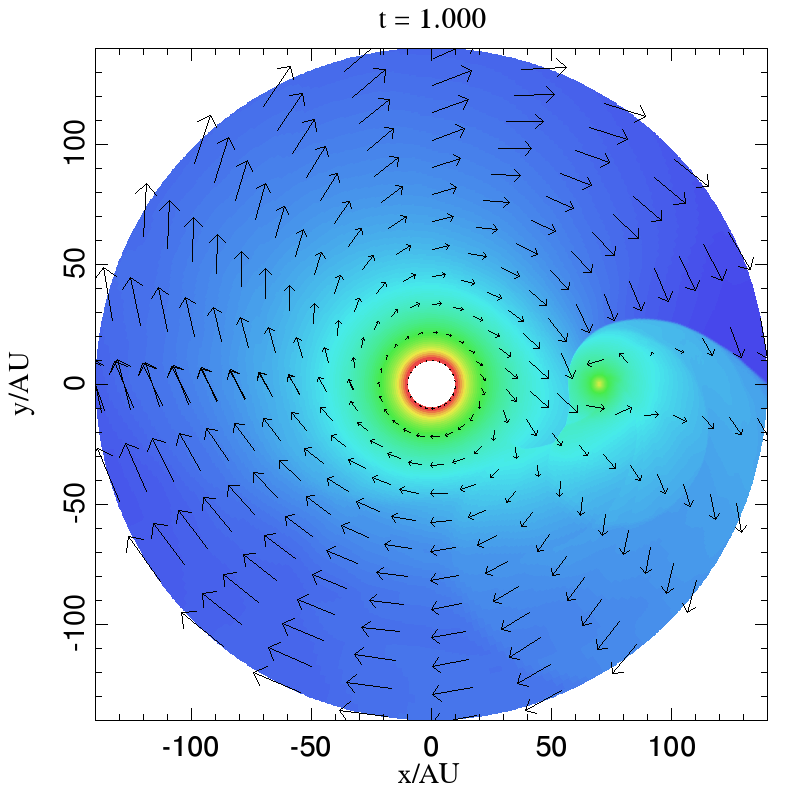}
   \caption{Density maps and velocity vectors
   in the orbital plane in logarithmic scale
   for a separation of $70 \mathrm{AU}$.
   Three snapshots are shown for times $t = 0.5, 1$ and $1.5$
   periods after the secondary is introduced.
     }
   \label{fig:mira}
\end{figure}

\begin{table}
   \caption[]{Models with separation 70 AU and mass ratio
   $q = 2$ for different dust formation radius.}
      \label{tbl:mira-models}
      \centering
      \begin{tabular}{ccccc}
            \hline \hline
            Model & Dust formation radius& Wind temperature & q & $h_B$ \\
            & (AU) & (K) &  \\
            \hline
            M-1 & 10 & $900$ & 2 & 0.1 \\
            M-2 & 20 & $490$ & 2 & 0.1 \\
            M-3 & 30 & $340$ & 2 & 0.1 \\
            M-4 & 40 & $300$ & 2 & 0.1 \\
            M-5 & 30 & $300$ & 2 & 0.05 \\
            \hline
      \end{tabular}
\end{table}

The mass accretion rates onto the secondary are of the
order of the predicted ones from Bondi-Hoyle
theory when the inner boundary of our grid is
less than $20 \mathrm{AU}$.
Here, we will focus on slow winds
with a relative speed at the position
of the accretor of $6-25 \kms$.
Mass accretion ratios for different 
values of the accretion radius $r_{\rm acc}$
inside the Hill radius
in Eq.~\ref{eq:acckley}
give very similar results after we reach
a stable configuration.
In our runs with 4 additional levels of refinement
around the accretor, the accretion ratio is
consistent with the lower resolution simulations.

\begin{figure}
    \includegraphics[width=0.5\textwidth]{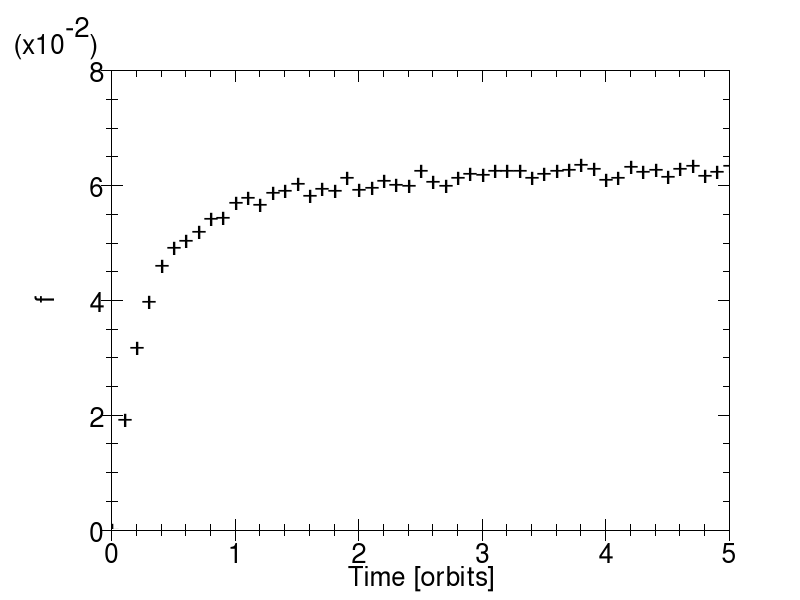}
    \caption{Accretion ratio, defined as the accretion
    rate divided by the mass-loss rate from the AGB
    star, as a function of time for
    our model M-5 (see Table~\ref{tbl:mira-models}).}
    \label{fig:accret}
\end{figure}

For a fast wind we expect
an accretion rate of the order of the Bondi-Hoyle
accretion \citep{1944MNRAS.104..273B}.
In this case we find a quasi-spherical wind morphology.
For wind acceleration radii between $30$ to $40
\mathrm{AU}$ the accretion rate is a few times the
Bondi-Hoyle rate with variations with an amplitude of
about 10\%.  The wind presents a broad range of density
contrasts in our computational domain.  Bow shocks are
formed close to the secondary in these models at late
times.
It has been suggested that UV radiation in symbiotic systems
is originated in those shocks \citep{1993MNRAS.265..946T}.
In Table~\ref{tbl:mira-models} we show the
parameters of our models that reproduce closely the
binary system Mira~AB for different wind acceleration
radii.
For typical mass loss
rates from an AGB star \citep{2000ApJ...545..945R}
we compute an accretion rate on
the companion of $\sim 10^{-7}-10^{-9}~\Msun~\yr$ via stream flow
accretion.
The X-ray luminosity of the accretor 
can be estimated using
\begin{equation}
    L_{\rm X} = \frac{G {\dot M}_{\rm acc}M_{\rm
    B}}{R_{\rm B}},
    \label{eq:luminosity}
\end{equation}
which corresponds to $\sim 3 \times 10^{30}-3 \times
10^{32}\,{\rm erg\,s}^{-1}$,
assuming a radius $R_{\rm B} = 0.02 \Rsun$ for the compact
object in our M-3 model.
In Fig.~\ref{fig:accret} we compare the accretion ratio in our models
with the Bondi-Hoyle accretion given by
Eq.~\ref{eq:accret}.
Table~\ref{tbl:accratio} shows accretion rates onto the compact object divided
by the mass loss rate of the AGB star as a function of time. 
Accretion ratios become quasi-stable after half of an orbit of
the system.
However, in the Bondi-Hoyle
formula the velocity is the upstream wind velocity at
infinity, while in our binary simulations we use the
relative wind velocity at the secondary position.
Moreover, the plane
parallel flow assumption in the Bondi-Hoyle
theory is not valid for our wind model at close
distances.

A small change in the wind parameters can lead to
changes of a few times in the accretion rate when an
accretion disk is formed around the secondary.  In
particular, it could explain the increase of extinction
in symbiotic binaries \citep{2008arXiv0804.4139G} as a change in
the accretion flow.
Those variations can occur in a
fraction of the orbital period and may also explain Nova
outbursts in objects such as \object{RS~Ophiuchi}
\citep{2008A&A...484L...9W}.

\begin{table}
   \caption[]{Accretion ratios as a function of time for
    the simulation shown in Fig.~\ref{fig:accret}.}
      \label{tbl:accratio}
      \centering
      \begin{tabular}{lcc}
            \hline \hline
            Time (orbits) &  $ \frac{\Mdot_\mathrm{acc}}{\Mdot_\mathrm{wind}}$ \\
            \hline
            0.1 & 0.0187499146615 \\
            0.2 & 0.032895734525 \\
            0.3 & 0.0405163717933 \\
            0.4 & 0.0457114513595 \\
            0.5 & 0.0486647436321 \\
            0.6 & 0.052120162122 \\
            0.7 & 0.0518595175058 \\
            0.8 & 0.0545558407777 \\
            0.9 & 0.0543526731365 \\
            1.0 & 0.0550009176397 \\
            1.1 & 0.0557649810422 \\
            1.2 & 0.0573139141097 \\
            1.3 & 0.0579923479442 \\
            1.4 & 0.0587394563781 \\
            1.5 & 0.0588350261856 \\
            1.6 & 0.058462177135 \\
            1.7 & 0.059570725981 \\
            1.8 & 0.0612462210888 \\
            1.9 & 0.0612788117294 \\
            2.0 & 0.0605904603547 \\
            \hline
      \end{tabular}
\end{table}

\begin{figure}
   \includegraphics[width=0.5\textwidth]{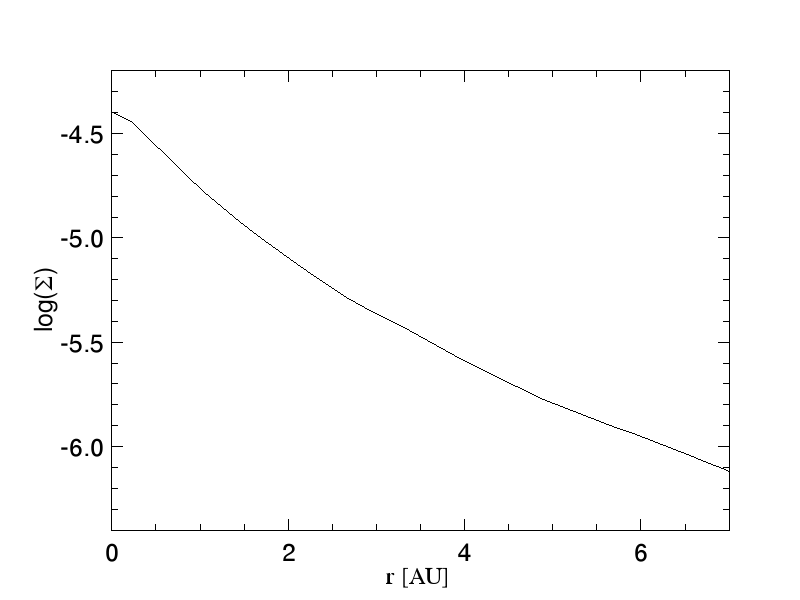}
   \caption{Averaged radial density profile of the disk
   around the accreting secondary
   in logarithmic scale in computational units after 1 orbit
   for the model shown in Fig.~\ref{fig:mira}.
     The distance in the horizontal axis is given in AU
   from the center of the secondary.
   }
   \label{fig:disk}
\end{figure}

\begin{figure}
   \includegraphics[width=0.5\textwidth]{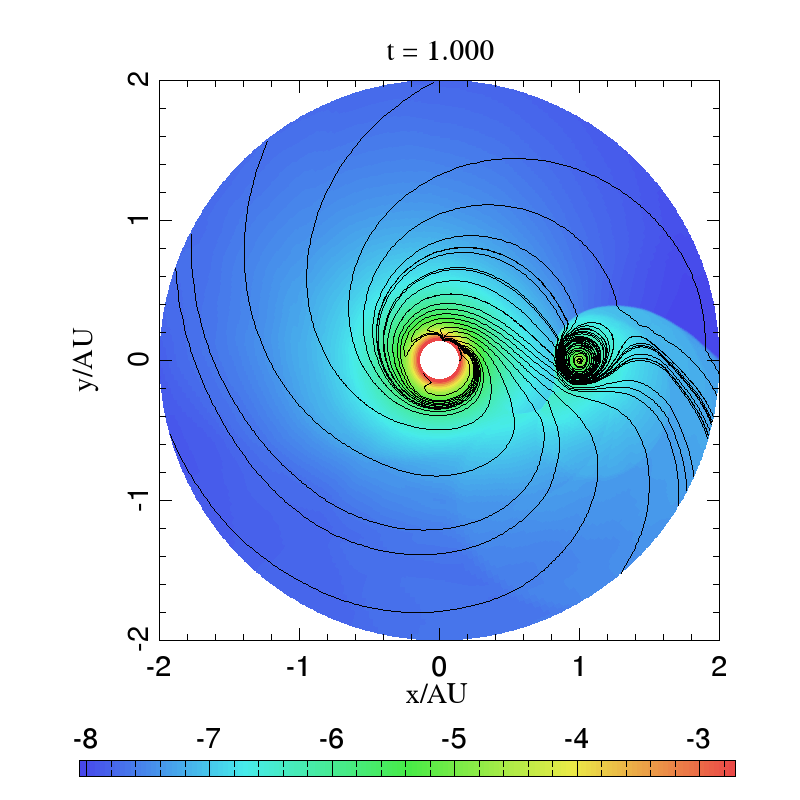}
   \caption{Streamlines in the corotating frame after 1 orbit
   overplotted on the
   density contours in logarithmic scale
   for the same model as shown in
   in Fig.~\ref{fig:mira}.
   The flowlines around the secondary are eccentric.
   }
   \label{fig:stream}
\end{figure}

In Fig.~\ref{fig:disk} the azimuthally averaged density
distribution close to the accreting star for model M-3 is shown.
The density decreases exponentially up to 5 AU in a slightly
eccentric disk and is likely to be optically thick.
This is consistent with the size of the disk around
Mira~B estimated by \citet{2007ApJ...662..651I}.
We show the flowlines after 1 orbital
period in Fig.~\ref{fig:stream}. The flowlines are
calculated integrating the equation of motion from the
line joining the stars using positive and negative
time-steps.  When the wind speed is of the order of the
escape velocity from the AGB star the flow is focused by
the gravitational pull from the secondary and the
accretion rate is enhanced by a factor of a few with
respect to standard Bondi-Hoyle accretion.

\begin{figure}
   \includegraphics[width=0.5\textwidth]{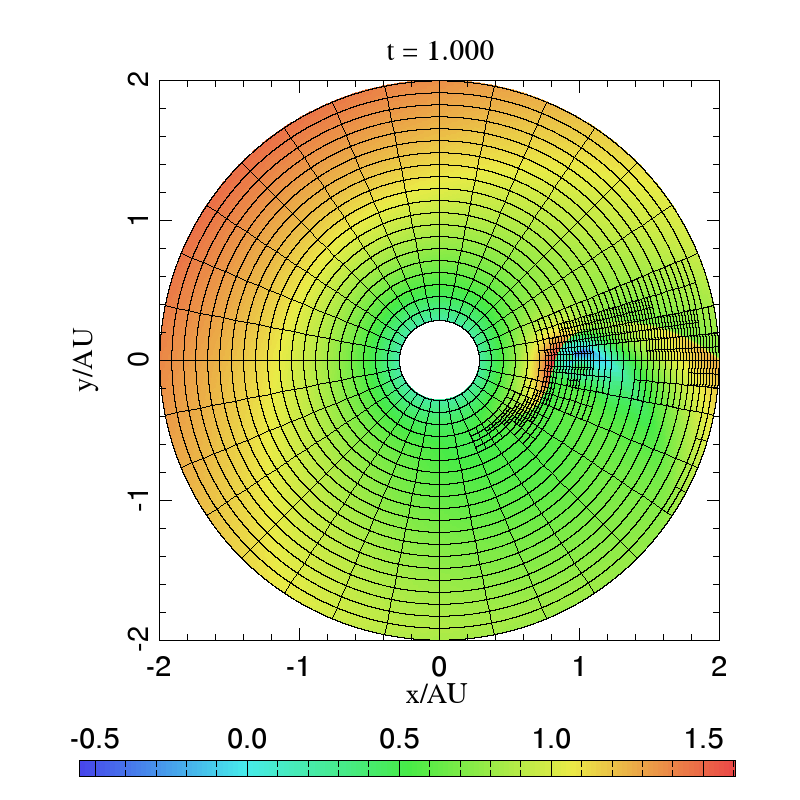}
   \caption{Radial velocity contours in the corotating frame after 0.5 orbits
   with grid structure using 3 levels of refinement. Each block has
   $8\times8$ computational zones.
   }
   \label{fig:blocks}
\end{figure}

We show the radial velocity map for a model with wind
temperature $10^3$ K and inner boundary at 20 AU with
overplotted grid structure in Fig.~\ref{fig:blocks}.
The refinement criterion used by {\sc PARAMESH} is based
on the second derivative normalized by the gradient over
one cell \citep{27908}.  The blocks are approximately
square close to the secondary and contain $8\times 8$
computational cells.

\begin{figure}
   \includegraphics[width=0.5\textwidth]{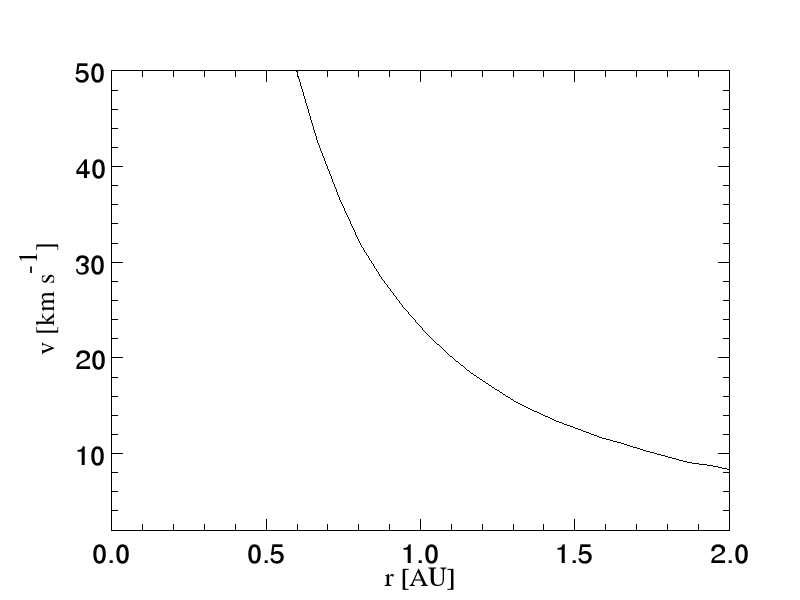}
   \caption{Averaged orbital velocity profile of the accretion disk around the
   secondary for the model shown in Fig.~\ref{fig:mira}.
     The distance in the horizontal axis is given in AU
   from the center of the secondary.
   }
   \label{fig:disk_vel}
\end{figure}

The orbital speed in the disk is close to Keplerian
with a steeper radial profile.  In
Fig.~\ref{fig:disk_vel}, we plot the averaged azimuthal
velocity around the secondary.

\begin{figure}
   \includegraphics[width=0.4\textwidth]{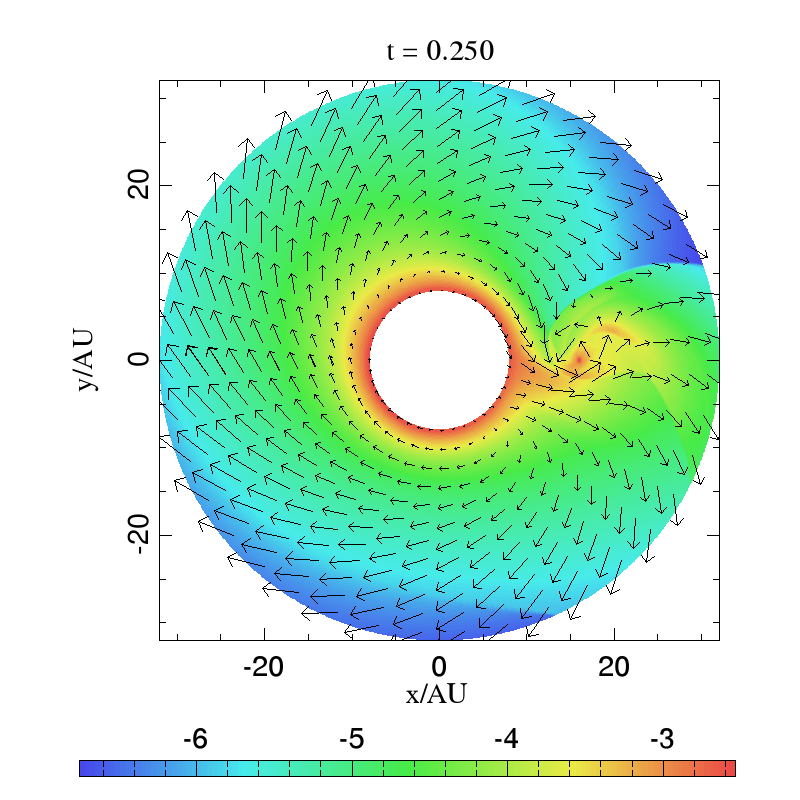}
   \includegraphics[width=0.4\textwidth]{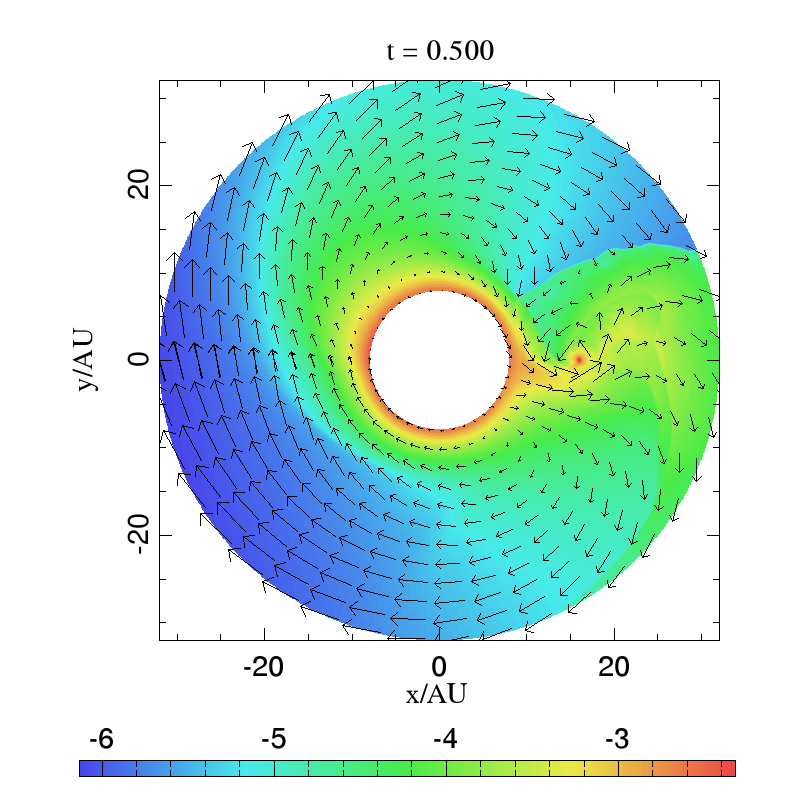}
   \includegraphics[width=0.4\textwidth]{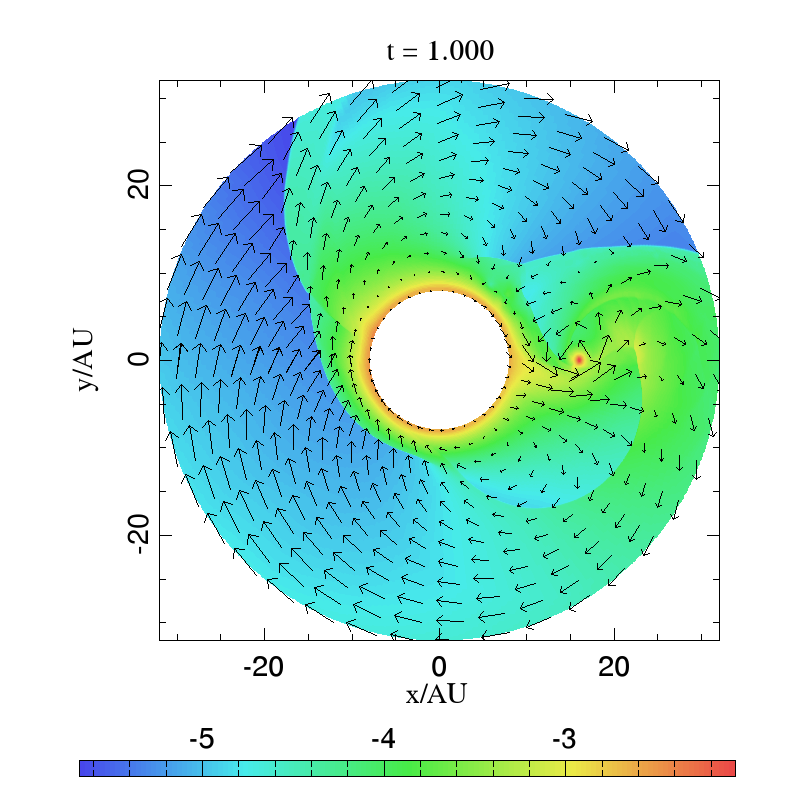}
   \caption{Density contours in the orbital plane
   with overplotted velocity vectors
   in the corotating frame
   for a system with separation of $16 \mathrm{AU}$
     after 0.25, 0.5 and 1 orbital periods.}
   \label{fig:Raqu}
\end{figure}

\subsubsection{Influence of the wind temperature}

We tried several models to check the dependence of
the wind parameters on the formation of a stream flow
between the stars.  The dust is believed to be condensed
at a temperature of $\sim 1500 {\rm K}$ which corresponds to a
distance of 8 AU from the center of the primary
\citep{1997MNRAS.287..799I}.
When the wind acceleration radius is of the order 50\% of the
Hill radius a density enhancement is formed focusing
10\% of the slow wind inside the Roche lobe.
The sound speeds in our models are in the range $3-10 \kms$
which are of the same order of the wind speed relative
to the accretor.
When the average velocity of the wind at the inner Lagrangian
point is larger than the orbital velocity of the system,
the stream does not form and the accretion ratio is
close to Bondi-Hoyle estimate.

In Fig.~\ref{fig:miratime}, we show the density contours
for a binary separation of $70 \mathrm{AU}$ for several
values of the dust formation radius where the wind is
accelerated.  In these simulations we include the full
gravity of the primary.

For a dust formation radius of $5 \mathrm{AU}$ the wind
geometry is almost spherical. A bow shock forms as
the wind is decelerated by the gravity from the
companion and the
accretion ratio is close to the Bondi-Hoyle value. 
A small accretion disk is formed around the
secondary.
As we move the dust formation radius outwards the
temperature for dust condensation decreases according to
Fig~\ref{fig:wind_temp}. When we place the dust formation
radius at $10 \mathrm{AU}$ the flow structure is clearly
elongated and the accretion rates are a few times larger
than the Bondi-Hoyle theoretical estimate.  This is
consistent with an agglomeration of dust particles at
several stellar radii above the photosphere in AGB
stars in order to have focused wind RLOF-like mass transfer
via stream flow in
wide symbiotic systems such as Mira~AB
\citep{2005ApJ...623L.137K}.

For the cases where the dust
formation is $\le 20 \mathrm{AU}$ the wind velocity is
lower than the escape velocity in our model.  The wind opposite from
the companion is not able to overcame the gravity and
forms a bow shock.  This does not affect the formation
of a tidal stream and our calculation of the mass
accretion rate onto the secondary.  The escape velocity
from the primary is $\sim 10 \kms$ at the dust formation
location in our model M-2 (second column in
Fig.~\ref{fig:miratime}).
In the slow wind case, numerical simulations 
show intricate flow patterns, therefore it 
is not possible to apply directly the results
from Bondi-Hoyle accretion theory.

In our model M-5,
we have changed the temperature profile
around the secondary
The accretion ratios averaged over one orbital period
agree within 5\% in our models M-3 and M-5.

\subsubsection{Modifying $\gamma$}
\label{sec:resultsgamma}

We have varied the polytropic index between
$\gamma =1-5/3$.
The density distributions are 
similar within 10\% for the simulations
with $q=2$, $d=70AU$ and
$\gamma =1,1.2,5/3$.
The variation of the polytropic index
intends to bracket more realistic cases
that include radiative losses in the
equation of state.
In the adiabatic simulations the accretion ratio
varies between $8-10$\% and is several
times larger than in the Bondi-Hoyle scenario.

\subsubsection{Variation of the orbital parameters}

The orbital velocity in our models varies between
$5-10 \kms$ while
the wind velocity at the surface of the AGB.
is of the order $2-10 \kms$.

In Fig.~\ref{fig:Raqu} the density contours of a system
with separation of $16 \mathrm{AU}$ and mass-losing
surface located at $8 \mathrm{AU}$ from the origin is
shown as a function of time.  The central star has a
mass of $1.2 \Msun$ and an accreting companion of
$0.6 \Msun$ orbits on a fixed circular orbit.
An accretion disk is observed around the secondary with
Keplerian velocity profile and a narrow stream forms
between the stars.  The accretion flow focused onto the
secondary is about 15\% of the unperturbed wind mass
loss which is several times more efficient than the
models for binary separation of $70 \mathrm{AU}$.
Mass is removed from the accretion disk
after each time-step to avoid the accumulation of mass
and formation of strong shocks according to
Eq.~\ref{eq:acckley} \citep{2002A&A...387..550G}.

\subsubsection{Convergence tests}
\label{sec:conv}

We have rerun simulation M-3 at a twice the linear
resolution of our reference simulations with three
levels of refinement, and performed an additional run at
the same base resolution with four additional levels of
refinement.
In our high resolution runs, the gravitational softening 
$\epsilon$ is half the size of the gravitational softening
in the low resolution runs.
The density distributions in the higher resolution
simulations agree with our
previous results within 5\% with larger oscillations.
Therefore, we conclude that the resolution
of our simulations is acceptable and does not affect our
results.

\section{Conclusions}
\label{sec:discuss}

We have constructed a hydrodynamical time-dependent model of a
wind from an AGB star which is gravitationally focused
by a close companion.  The model has several free
parameters.  The mass ratio $q$ of the system, the
binary separation $d$ and the eccentricity $e$ characterize the binary
configuration.  The effective temperature on the surface
of the mass losing star, the radius of the AGB star
where the dust is formed and the mass loss describe the
properties of the wind from the mass-losing star.  We
have improved on previous hydrodynamic simulations by
using an equation of state where the temperature
distribution depends on the distance from the primary
star and the accreting companion.

For each binary model we integrate the equations of
hydrodynamics until the density distribution of the
gravitationally focused wind reaches a stationary state.
The general effect of the secondary is to deflect the
AGB wind toward the orbital plane, and as a result the
relaxed density distributions present accretion disks of various
characteristics.
Accretion via a gravitationally focused wind has implications for a broad range
of symbiotic and wide binary systems and can explain the formation of
chemically peculiar stars.
A tidal stream is observed in our runs
for binary separations up to $\sim 70 AU$ for slow
winds and nearly isothermal flows.
Wind focusing weakens with increasing orbital separation because of the
increasing ratio of the wind velocity at the secondary's position to the
orbital velocity of the system.

All our models develop a high density region in the form
of a Keplerian thin disk around the secondary with a
mass of about $10^{-4} \Msun$ and an exponentially
decreasing density profile.
The temperature profile around the accretor was fixed
in our simulations and corresponds to a disk scale height of $\sim0.05$.
Winds from evolved stars in binary systems
show complex flow patterns and are far from being
spherically symmetric.
A thin accretion disk
around Mira~B has been inferred from spectroscopic UV
observations
\citep[e.g.][]{1985ApJ...297..275R,1997ApJ...482L.175K} and
interferometric imaging
\citep{2007ApJ...662..651I}.  The formation
of a stream flow is strongly dependent on the wind
temperature at the dust formation radius which
determines the wind velocity at the inner Lagrangian
point.  For the winds parameters from a AGB star (Mira type)
we obtain stable disks with masses of the order
$10^{-4}$ of the stellar mass for a range of orbital
separations.  The accretion rate onto the secondary is
almost constant once the accretion disk is formed and we
reach a quasi-stable state.  We conclude that the main
parameter that determines the formation of a stream and
enhanced accretion rates compared to Bondi-Hoyle's rates
is the wind velocity at the dust formation radius.
Nevertheless, the Bondi-Hoyle estimate does
not include pressure forces
and is not directly applicable to our models
of wind accretion in symbiotic binaries.

The presence of a long-lived disk and stream flow is sensitive
to the wind parameters, in particular of the dust
acceleration radius where we place the inner boundary of
our grid.  In case that the wind parameters change with
time the accretion rates can be expected to change in
relatively short time-scales due to rapid
changes in the flow patterns and the formation of a
stream flow.  This can explain the variability in
accretion rate and transitions from a quiescent to an
active state in binary systems
\citep{2002ARep...46.1022B}.
As the secondary increases in mass
there are several possible outcomes for the system.
The angular momentum loss
in symbiotic binaries may be large enough 
to imply a shrinking of the orbit and a possible merger
that can lead to the formation of Barium stars or Type
Ia supernovae \citep{1996MNRAS.280.1264T,2005A&A...441..589J}.

The massive wind of an AGB star can be considerably deflected towards the
orbital plane by the gravitational interaction with a companion star.
The X-ray luminosity powered by accretion onto the mass
gainer can photoionize the wind and reduce its ability
to be accelerated by radiation pressure.  The lower
radiative driving can result in a reduced wind velocity
near the companion, which further increases the
the accretion efficiency of the companion.  In the case of rapid
accretion rate changes a symbiotic binary may undergo
outbursts such as those observed in Mira~AB
\citep{2005ApJ...623L.137K}.
A detailed comparison of the results from
the numerical models with multi-wavelength observations
of Mira~AB will be presented elsewhere.

We have adopted a simple a simple prescription for the
equation of state and wind acceleration mechanism.  In
the future our calculations can be improved by using a
more detailed energy balance for the gas going beyond
the isothermal approximation including radiation cooling
and heating due to shock formation.
Radiative feedback is also expected to be important
in the context of symbiotic systems and 
X-ray binaries~\citep{2004MNRAS.349..678E}.
We plan to perform
3-dimensional simulations to compare our numerical
results with high resolution X-ray observations of the
Mira~AB system.

\begin{acknowledgements}
MdVB was supported by a SAO predoctoral fellowship
during the course of this work.
MK is a member of the Chandra X-ray Center, which is
operated by the Smithsonian Astrophysical Observatory
under NASA contract NAS8-03060.
The \flash{} code used
in this work is developed in part by the U.S. Department
of Energy under Grant No. B523820 to the Center for
Astrophysical Thermonuclear Flashes at the University of
Chicago.  Some of the simulations reported here were
performed at the Center for Parallel Astrophysical
Computing operated by the Institute for Theory and
Computation at the Harvard-Smithsonian Center for
Astrophysics.  MdVB thanks P.\ Artymowicz for valuable
discussions.
We would like to thank the anonymous referee
for helpful comments and suggestions.
\end{acknowledgements}

\bibliographystyle{hapj}
\bibliography{references,ref,arxiv}

\end{document}